%% file: JUNO-HighPrecisionReactorNuOsc.tex
\documentclass[a4paper,11pt]{article}
\usepackage[left=25mm,right=25mm,top=20mm,bottom=20mm]{geometry}

\usepackage[T1]{fontenc} 
\usepackage{bm}
\usepackage{subfigure}
\usepackage[affil-it]{authblk}
\usepackage{tabularx}                        	%
\usepackage{graphicx}                        	%
\usepackage{amsmath}                        	%
\usepackage{amssymb}                        	%
\usepackage{float} 				
\floatstyle{plaintop}
\restylefloat{table}
\usepackage{caption}
\usepackage{units}				%
\usepackage{siunitx}
\usepackage{gensymb}
\usepackage[normalem]{ulem}
\usepackage{xcolor}				%
\usepackage{cite}

\newcommand{\dmN}{$\Delta m^2_{21}$}
\newcommand{\Dm}{$\Delta m^2_{31}$}

\title{Sub-percent Precision Measurement of Neutrino Oscillation Parameters with JUNO}

\input{fullname_211222}

\begin{document}
\maketitle

\begin{abstract}
JUNO is a multi-purpose neutrino observatory under construction in the south of China. This publication presents new sensitivity estimates for the measurement of the $\Delta m^2_{31}$, $\Delta m^2_{21}$, $\sin^2 \theta_{12}$, and $\sin^2 \theta_{13}$ oscillation parameters using reactor antineutrinos, which is one of the primary physics goals of the experiment. The sensitivities are obtained using the best knowledge available to date on the location and overburden of the experimental site, the nuclear reactors in the surrounding area and beyond, the detector response uncertainties, and the reactor antineutrino spectral shape constraints expected from the TAO satellite detector. It is found that the $\Delta m^2_{31}$, $\Delta m^2_{21}$, and $\sin^2 \theta_{12}$ oscillation parameters will be determined to better than 0.5\% precision in six years of data collection, which represents approximately an order of magnitude improvement over existing constraints.

\end{abstract}

\section{Introduction}

Neutrinos have provided us with the first direct evidence of physics beyond the Standard Model (SM)
of elementary particles, and the study of their properties stands as one of the most active directions within particle physics.
First detected by Reines and Cowan in 1956~\cite{Cowan:1992xc}, these particles were discovered {to oscillate} roughly four decades later~\cite{Kajita:2016cak,McDonald:2016ixn}, an unambiguous sign that {they are massive and that the SM needs to be modified.}

The phenomenology of neutrino oscillations granted elegant solutions to
the solar neutrino~\cite{SNO:2002tuh}
and atmospheric neutrino~\cite{Super-Kamiokande:1998kpq} {anomalies} through the transformation of electron and muon neutrinos into other neutrino flavors, respectively.
To date,
almost all neutrino data collected with accelerator, solar, atmospheric, and reactor neutrinos~\cite{Zyla:2020zbs}
can be explained within the standard three-neutrino oscillation paradigm
that will be described in the next section.
It is essential, however, that the accuracy of {this} paradigm {be} scrutinized with increasing precision to inform
future experiments, provide important constraints to model building, and {probe} for other physics beyond the SM.

\subsection{Neutrino Masses and Mixing}

{In the standard three neutrino flavor scheme,}
neutrino oscillations imply that
{there exist three distinct neutrino mass eigenstates
possessing definite} neutrino masses, $m_i$ ($i$=1, 2, 3), {which} are non-degenerate, namely, $m_i \ne m_j$ for $i\ne j$.
This in turn implies that at least two neutrino species
must be massive.
In such a non-degenerate neutrino mass spectrum,
each known flavor eigenstate, (${\nu_e}$, ${\nu_\mu}$, ${\nu_\tau}$),
linked to three respective charged leptons ($e$, $\mu$, $\tau$) via the charged current interactions,
can be regarded as a non-trivial mixture of the neutrino mass eigenstates as
\begin{eqnarray}
\left(
\begin{matrix}
\nu_e  \cr
\nu_\mu  \cr
\nu_\tau  \cr
\end{matrix}
\right)
=
   U_\text{\tiny PMNS}
\left(
\begin{matrix}
\nu_1  \cr
\nu_2  \cr
\nu_3  \cr
\end{matrix}
\right),
\end{eqnarray}
where $\nu_i$ ($i$=1, 2, 3) denote the mass eigenstates,
and $U_\text{\tiny PMNS}$ is the so called
{\it Pontecorvo-Maki-Nakagawa-Sakata} (PMNS)~\cite{Pontecorvo:1967fh,Maki:1962mu} matrix,
a $3\times 3$ unitary matrix describing neutrino mixing.
The first mass eigenstate $\nu_1$ is defined as the one with the largest portion of the electron flavor eigenstate ${\nu_e}$.
The mixing matrix for antineutrinos is a complex conjugate of
the one for neutrinos, $U_\text{\tiny PMNS} \to U_\text{\tiny PMNS}^\ast$.
The standard parametrization of the PMNS matrix is given by~\cite{Zyla:2020zbs},
\begin{eqnarray}
   U_\text{\tiny PMNS} =
\left(
 \begin{matrix}
 1  &  0       & 0       \cr
 0  &  c_{23}  & s_{23}  \cr
 0  & -s_{23}  & c_{23}  \cr
\end{matrix}
\right)
\left(
\begin{matrix}
 c_{13} & 0  &  s_{13}e^{-i\delta_{\text{\tiny CP}}} \cr
 0  & 1 & 0 \cr -s_{13} e^{i\delta_{\text{\tiny CP}}} & 0 & c_{13} \cr
\end{matrix}
\right)
\left(
\begin{matrix}
c_{12} & s_{12} & 0 \cr
-s_{12} & c_{12} & 0 \cr
0  & 0 & 1 \cr
\end{matrix}
\right)
\left(
\begin{matrix}
e^{i\eta_1} & 0 & 0 \cr
0 & e^{i\eta_2}  & 0 \cr
0  & 0 & 1 \cr
\end{matrix}
\right),
\label{eq:U}
\end{eqnarray}
where the notation
$c_{ij} \equiv \cos \theta_{ij}$ and
$s_{ij} \equiv \sin \theta_{ij}$ is used,
and $\delta_{\text{\tiny CP}}$ is often called the Dirac CP phase.
Here, $\eta_i\;(i=1,2)$ are the Majorana CP phases, which are physical only if neutrinos are Majorana-type particles
but play no role in
neutrino oscillations~\cite{Bilenky:1980cx}.

{As shown later, a total of six parameters are needed to fully describe neutrino oscillations, namely, three mixing angles, one {Dirac} CP phase, and two independent mass squared differences. The latter characterize the degree of non-degeneracy of neutrino masses and are defined as
\begin{eqnarray}
\Delta m^2_{ij} \equiv m_i^2-m_j^2 \ \ \ (i,j = 1,2,3,\, i > j).
\label{eq:Dm2}
\end{eqnarray}
As will be discussed in detail in this publication, among these six parameters,
{the Jiangmen Underground Neutrino Observatory (JUNO)}
can significantly improve the precision of $\Delta m^2_{31}$ (or equivalently $\Delta m^2_{32}$), $\Delta m^2_{21}$, and $\sin^2 \theta_{12}$.
In addition, JUNO can measure ${\sin^2 \theta_{13}}$ but with less precision {than current reactor experiments~\cite{Adey:2018zwh,DoubleChooz:2019qbj,Bak:2018ydk}},
and is insensitive to $\sin^2 \theta_{23}$ and $\delta_{\text{\tiny CP}}$.
}

\subsection{Today's Knowledge on Oscillation Parameters}

Table~\ref{tab:NuToday} presents the current precision of {the} four mixing parameters within the reach of JUNO as obtained from {the 2020 Review of Particle Physics} \cite{Zyla:2020zbs}, which is referred to as PDG2020. {The mass ordering (MO) of neutrinos is still unknown, meaning that the sign of $\Delta m^2_{32}$ can be positive or negative. The normal mass ordering (NMO) corresponds to $\Delta m^2_{32} > 0$, and the inverted mass ordering (IMO) to $\Delta m^2_{32} < 0$.}
The most recent global analyses~\cite{Esteban:2020cvm,deSalas:2020pgw,Capozzi:2021fjo}
yield estimations that are consistent with the values shown in Table~\ref{tab:NuToday}. Current precision on most oscillation parameters is in the order of a few percent.

\begin{table}[h!]
\begin{center}
\begin{tabular}{lcc}
\hline
& PDG2020 & Relative Uncertainty (1$\sigma$)  \\
\hline
$\Delta m^2_{32}$ (NMO) & (2.453$\pm$0.034) $\times 10^{-3}$ eV$^2$  & 1.4\%
\\
$\Delta m^2_{32}$ (IMO) & $-(2.546\pm$0.037) $\times 10^{-3}$ eV$^2$  & 1.5\%
\\
$\Delta m^2_{21}$ & (7.53$\pm$0.18) $\times 10^{-5}$ eV$^2$  &2.4\%
\\
$\sin^2 \theta_{12}$  & 0.307$\pm$0.013  & 4.2\%
\\
$\sin^2 \theta_{13}$  & 0.0218$\pm$0.0007  & 3.2\%
\\
\hline
\end{tabular}
\caption{
  \small {
  Today's best knowledge of neutrino oscillation parameters  within the reach of JUNO and their 1$\sigma$ uncertainties, as reported in the PDG2020~\cite{Zyla:2020zbs}. The relative uncertainties (in \%) are indicated in the last column. NMO (IMO) implies normal mass ordering (inverted mass ordering).
}      }
\label{tab:NuToday}%
\end{center}
\end{table}%

\subsection{New Knowledge to be Provided by JUNO}
One of the main goals of JUNO~\cite{An:2015jdp, JUNO:2022hxd} is to determine
{the neutrino MO.}
This can be done by precisely measuring the interference in the reactor antineutrino
oscillation probability driven by two independent mass squared differences,
$\Delta m^2_{31}$ and $\Delta m^2_{32}$, as originally considered in Ref.~\cite{Petcov:2001sy}.
{Additional details on JUNO's MO determination can be found in Ref.~\cite{An:2015jdp}, and an updated estimate is under preparation.}
{JUNO's physics program also includes studies of neutrinos from the Sun~\cite{Abusleme:2020zyc}, the atmosphere~\cite{JUNO:2021tll}, supernovae~\cite{Lu:2016ipr}, and planet Earth~\cite{An:2015jdp}, as well as explorations of physics beyond the SM~\cite{An:2015jdp}.}

{JUNO's measurement of the oscillated reactor antineutrino spectrum at $\sim$52.5 km will also enable an independent determination of the $\Delta m^2_{31}$, $\Delta m^2_{21}$, $\sin^2 \theta_{12}$, and $\sin^2 \theta_{13}$ oscillation parameters, which is the focus of this publication. Of these, the first three will be determined to significantly better than $1\%$, inaugurating a new era of precision in neutrino oscillation measurements~\cite{An:2015jdp,DUNE:2020ypp,Hyper-Kamiokande:2018ofw}.}
Such extraordinary precision is expected to have a vast impact across different research fields {including particle physics, astrophysics, and cosmology}. For instance, it will enable more stringent tests of the standard 3 flavor neutrino mixing picture, such as probing the unitarity of the PMNS matrix~\cite{Antusch:2006vwa,Parke:2015goa,Fong:2016yyh,Blennow:2016jkn,Li:2018jgd,Ellis:2020hus}, {with the potential to discover physics beyond the SM.}
It {will also have} important implications for other experimental efforts,
{for example} by reducing the parameter space in the search for leptonic CP violation~\cite{DUNE:2015lol,Hyper-KamiokandeProto-:2015xww} and
neutrinoless double beta decay~\cite{Dueck:2011hu,Ge:2015bfa,Cao:2019hli}. The precise {knowledge} of the leptonic mixing matrix may reveal its most fundamental structure and provide important clues for identifying
the theoretical mechanisms behind neutrino mass and mixing generation~\cite{King:2019gif}.
Finally, the new precision {will allow using} neutrinos as a more reliable tool or messenger to probe the deep interiors of astrophysical objects such as the Sun, supernovae, and planet Earth.


{Important updates are made compared to the previous estimate~\cite{An:2015jdp} of JUNO's sensitivity to the $\Delta m^2_{31}$, $\Delta m^2_{21}$, $\sin^2 \theta_{12}$, and $\sin^2 \theta_{13}$ oscillation parameters. Only eight nuclear reactors at $52.5$~km are considered instead of the ten envisioned when the experiment was first conceived. Likewise, realistic IBD selection and muon veto efficiencies obtained with the state of the art simulation software, as well as realistic assumptions on the detector performance 
and the final information about the location and overburden of the experimental site, are used. Finally, the expected constraints on the reactor antineutrino spectral shape from the satellite Taishan Antineutrino Observatory (TAO)~\cite{Abusleme:2020bzt:TAO_CDR} are employed. All these inputs are described in detail in the following sections.}

The remainder of this publication is organised as follows.
Sections 2 and 3 introduce the neutrino oscillation framework and the experimental setup of JUNO, respectively.
Section 4 provides the specifics related to reactor antineutrino detection and selection. Section 5 describes the methodology used to perform the oscillation analysis and parameter extraction, together with the main results. Section 6 is dedicated to the conclusions.

\section{Neutrino Oscillation Framework for JUNO}
\label{sec:osc_framework}

In the presence of non-degenerate neutrino masses and non-trivial mixing,
neutrinos and antineutrinos undergo flavor oscillations when they propagate in vacuum or in matter.
In this section, we present the neutrino oscillation framework used in the rest of this {work}.

 \subsection{Neutrino Oscillation in Vacuum}

The general $U_\text{\tiny PMNS}$-parameterization-independent expression
of the $\nu_\alpha \to \nu_\beta$ neutrino oscillation probabilities
in vacuum for ultra-relativistic neutrinos is given by~\cite{Zyla:2020zbs}
\begin{eqnarray}
  {\cal P}(\nu_\alpha \to \nu_\beta)
   =  \delta_{\alpha \beta}
 & - & 4 \sum_{i > j}
  \Re \left( U_{\alpha i}^\ast U_{\alpha j} U_{\beta i} U_{\beta j}^\ast \right)
\sin^2 \left( \frac{\Delta m^2_{ij}L}{4E} \right) \nonumber \\
  && +  2 \sum_{i > j}
  \Im \left( U_{\alpha i}^\ast U_{\alpha j} U_{\beta i} U_{\beta j}^\ast \right)
\sin \left( {\frac{\Delta m^2_{ij}L}{2E}} \right),
\label{eq:osc-prob-vac-general}
\end{eqnarray}
where 
$U_{\alpha i}$, $U_{\beta j}$, $U_{\alpha j}$, $U_{\beta i}$ (with $\alpha,\beta=e, \mu, \tau$ and $i,j=1,2,3$ being the flavor and mass indices respectively) are PMNS matrix elements,
$L$ is the distance traveled by the neutrino, $E$ {is} the neutrino energy,
and
$\Delta m^2_{ij}$ are the mass squared differences defined in Eq.~\eqref{eq:Dm2}.
For antineutrinos,
the mixing matrix elements in Eq.~\eqref{eq:osc-prob-vac-general} {are} replaced by their complex conjugates, as mentioned before.

 In this {work}, we are particularly interested in the case where $\alpha = \beta =e$
in Eq.~\eqref{eq:osc-prob-vac-general}, which yields the survival probability of electron antineutrinos.
Due to CPT invariance, the neutrino survival probabilities
for neutrinos and antineutrinos {are identical and given by:}
\begin{eqnarray}
{\cal P}(\overline{\nu}^{}_e \to \overline{\nu}^{}_e)
&  =  & 
1 - \hspace{0.05cm} \sin^2 2\theta_{12}\, c_{13}^4\, \sin^2 \Delta_{21}
-  \sin^2 2\theta_{13}
\left( c_{12}^2\sin^2 \Delta_{31} + s_{12}^2 \sin^2 \Delta_{32}\right)
\nonumber \\
&  =  &
1 - \hspace{0.05cm} \sin^2 2\theta^{}_{12} c_{13}^4 \sin^2 \Delta_{21}
- \frac{1}{2}
\sin^2 2\theta^{}_{13} \left(\sin^2 \Delta_{31}
+ \sin^2 \Delta_{32}\right)   \hspace{0.5cm}
\nonumber \\
& & \hspace{0.3cm} - \hspace{0.1cm}
\frac{1}{2} \cos 2\theta^{}_{12} \sin^2 2\theta^{}_{13}
\sin \Delta_{21}
\sin (\Delta_{31} + \Delta_{32}),
\label{eq:P_ee_vac}
\end{eqnarray}
where $\Delta_{ij} \equiv \Delta m^2_{ij}L/(4E)$,
and the standard parametrization of the mixing matrix shown in Eq.~\eqref{eq:U} has been used.
Note that in the second and third lines of the above equation,
we have reformulated the survival probability in a manner that
different parts of the solar-dominated (second term), atmospheric-dominated (third term),
and MO-sensitive oscillations (fourth term), are factored out.
Note also that there is no dependence {on}
either $\sin^2 \theta_{23}$ or $\delta_{\text{\tiny CP}}$.

\subsection{Neutrino Oscillation in Matter}
\label{sec:NO_Matter}

{Even though matter effects are relatively small in JUNO compared to long-baseline oscillation experiments, it is necessary to account for them to extract the correct values of the mixing parameters.}
{In fact,} ignoring matter effects would lead {to biases in} $\Delta m^2_{21}$ and $\sin^2\theta_{12}$ of about $1$\% and 0.2\%~\cite{Li:2016txk}, {respectively}.
{A complete treatment of the impact of matter effects in JUNO can be found in Refs.~\cite{Li:2016txk,Capozzi:2013psa,Khan:2019doq}, and this Section offers only a brief synopsis of the main points.}

The effective Hamiltonian that is responsible for the
antineutrino propagation in matter~\cite{Wolfenstein:1977ue,Mikheyev:1985zog} is given by
\begin{eqnarray}
\widetilde{\cal H}^{}_{\rm eff}  = \frac{1}{2 E} \left[U
\begin{pmatrix} m^2_1 & 0 & 0 \cr 0 & m^2_2 & 0 \cr 0 & 0 & m^2_3
\cr \end{pmatrix} U^\dagger - \begin{pmatrix} A & 0 & 0 \cr 0 & 0 &
0 \cr 0 & 0 & 0 \cr \end{pmatrix} \right] = \frac{1}{2 E} \left[\widetilde{U}
\begin{pmatrix} \widetilde{m}^2_1 & 0 & 0 \cr 0 & \widetilde{m}^2_2
& 0 \cr 0 & 0 & \widetilde{m}^2_3 \cr \end{pmatrix}
\widetilde{U}^\dagger \right]\; ,
\end{eqnarray}
where $\widetilde{U}$ and $\widetilde{m}^{}_i$ stand, respectively, for the effective
neutrino mixing matrix and the $i$-th neutrino mass in matter.
The matter parameter $A$ can be expressed as
\begin{eqnarray}
  A=2\sqrt{2} \ G^{}_{\rm F} N^{}_e E\simeq 1.52 \times 10^{-4} ~{\rm eV}^2 \times Y_e^{}
 \left[ \frac{\rho}{{\rm g}\cdot{\rm cm}^{-3}}\right]\, \left[ \frac{E}{\rm GeV}\right],
\end{eqnarray}
where $Y_e^{} \simeq 0.5$ is the electron fraction per nucleon and
$\rho = (2.45\pm0.15) ~{\rm g}/{\rm cm}^3$ is the estimated average matter density with its associated uncertainty, obtained for JUNO by considering that the antineutrino trajectory passes through both the crust and the sediment of the Earth.
Note that the minus sign in front of $A$ denotes the charged-current matter potential of electron antineutrinos in matter.

In JUNO, where matter effects are relatively small, a constant matter density profile can be assumed and the survival probability can be written in an analogous form as in vacuum,
by simply replacing the mass eigenvalues and mixing angles used in Eq.~\eqref{eq:P_ee_vac} by those in matter, which are indicated
with a tilde placed over the corresponding quantities as
\begin{eqnarray}
{\cal P}(\overline{\nu}^{}_e \to \overline{\nu}^{}_e)
&  =  & 
1 - \hspace{0.05cm} \sin^2 2\widetilde\theta_{12}\, {\tilde c}_{13}^4\, \sin^2 \widetilde\Delta_{21}
-  \sin^2 2\widetilde\theta_{13}
\left( \tilde c_{12}^2\sin^2 \widetilde\Delta_{31} + \tilde s_{12}^2 \sin^2 \widetilde\Delta_{32}\right)
\nonumber \\
&  =  &
1 - \hspace{0.05cm} \sin^2 2\widetilde\theta^{}_{12} \tilde c_{13}^4 \sin^2 \widetilde\Delta_{21}
- \frac{1}{2}
\sin^2 2\widetilde \theta^{}_{13} \left(\sin^2 \widetilde\Delta_{31}
+ \sin^2 \tilde \Delta_{32}\right)   \hspace{0.5cm}
\nonumber \\
& & \hspace{0.3cm} - \hspace{0.1cm}
\frac{1}{2} \cos 2\widetilde \theta^{}_{12} \sin^2 2\widetilde \theta^{}_{13}
\sin \tilde \Delta_{21}
\sin (\widetilde \Delta_{31} + \widetilde \Delta_{32}),
\label{eq:P_ee_mat}
\end{eqnarray}
where
$\tilde c_{ij} \equiv \cos \widetilde \theta_{ij}$,
$\tilde s_{ij} \equiv \sin \widetilde \theta_{ij}$,
with $\widetilde \theta_{ij}\, (i,j = 1,2,3,\, i<j)$ being the effective mixing
angles in $\widetilde U$
with the standard parametrization as in Eq.~\eqref{eq:U}.

An exact calculation of the survival probability can be obtained by numerical derivations of the eigenvalues and eigenvectors of $\widetilde{\cal H}^{}_{\rm eff}$, in principle, for an arbitrary matter density profile along the neutrino trajectory.
On the other hand, there are also several analytical approximations based on different expansion methods performed
under the constant matter density assumption.
For some approximated analytic formulae of
${\cal P}(\overline{\nu}^{}_e \to \overline{\nu}^{}_e)$ and effective mixing parameters in matter for JUNO, please refer to, for example,
Refs.~\cite{Li:2016txk,Capozzi:2013psa,Khan:2019doq}.
In the following sensitivity studies, both the exact calculation and analytical approximations were employed and found to produce consistent results.

\section{JUNO Experimental Setup}

\begin{figure}[!h]
  \centering
     \includegraphics[width=0.7\textwidth]{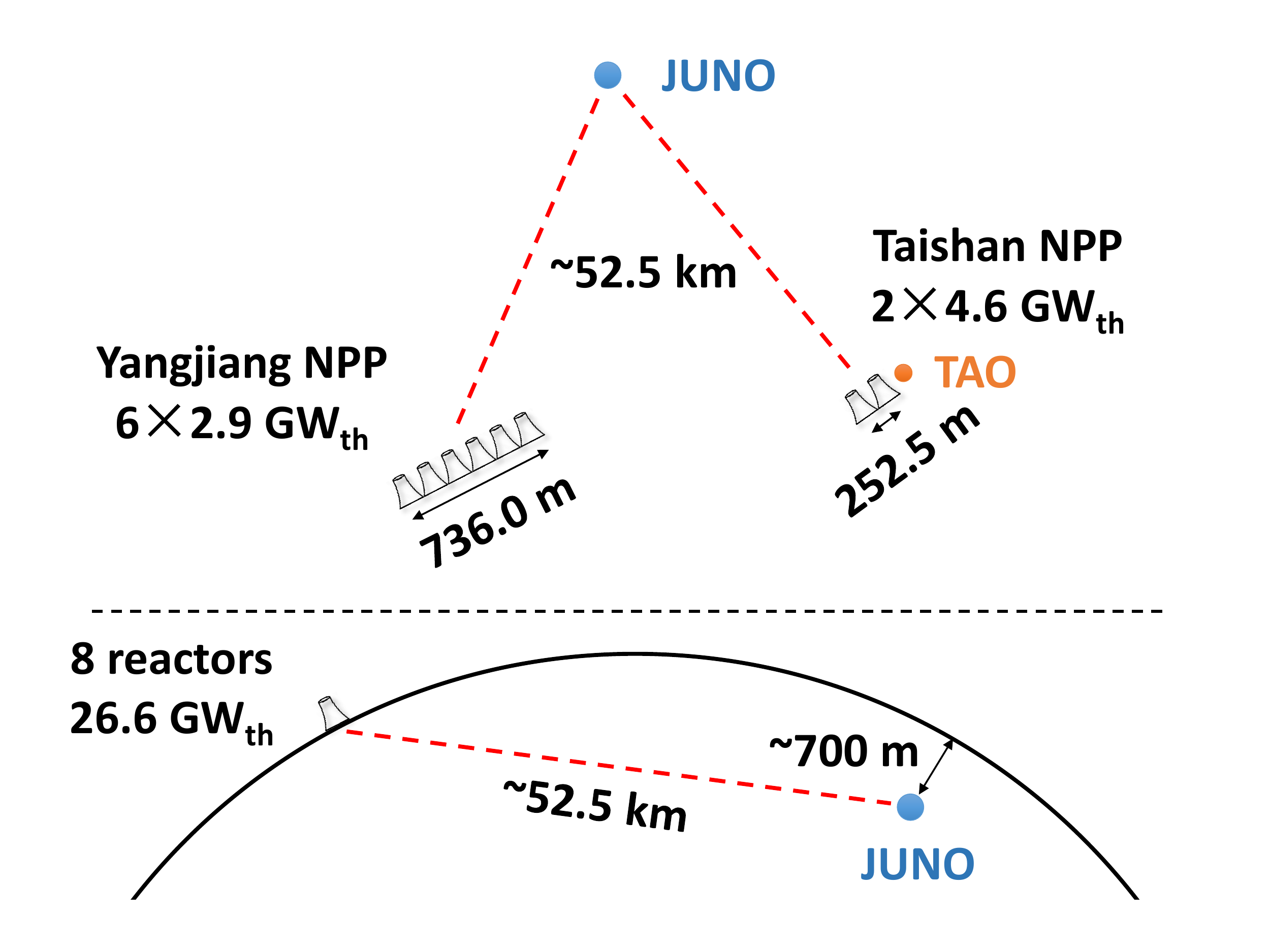}
  \caption{Setup of the JUNO experiment. The main 20~kton JUNO detector, indicated in blue, is located in an experimental cavern at a depth of about 700~m with respect to the surface and 650~m of overburden (1800~m.w.e), at a baseline of $\sim$52.5~km from six 2.9~GW$_\mathrm{th}$ reactor cores in the Yangjiang NPP and two 4.6~GW$_\mathrm{th}$ cores in the Taishan NPP. The 2.8~ton TAO detector, indicated in orange, is located about 30~m away from one of the Taishan reactor cores.}
  \label{fig:setup}
\end{figure}

JUNO is a multipurpose experiment currently under construction in Southern China that will use a 20 kton liquid scintillator target to study neutrinos from a variety of natural sources as well as from nuclear reactors. Most reactor antineutrinos in JUNO will originate from 2 and 6 cores in the Taishan and Yangjiang nuclear power plants (NPPs), respectively. Both plants are located at a baseline of about 52.5\,km, which was optimized for the best sensitivity to the neutrino MO and have a combined nominal thermal power of 26.6~GW$_{\rm th}$. Knowledge of the unoscillated reactor antineutrino spectrum shape is important for JUNO, so a dedicated small satellite detector~\cite{Abusleme:2020bzt:TAO_CDR}, called TAO, will be placed at about 30~m from one of the Taishan reactors to precisely measure it, serving as a data-driven input to constrain the spectra of the other cores. A schematic illustrating the location of both JUNO and TAO is shown in Fig.~\ref{fig:setup}. The experiment's main detector and the reactors considered in the analysis are described in detail in the following Subsections.

\begin{figure}[!h]
  \centering
  \includegraphics[width=0.7\textwidth]{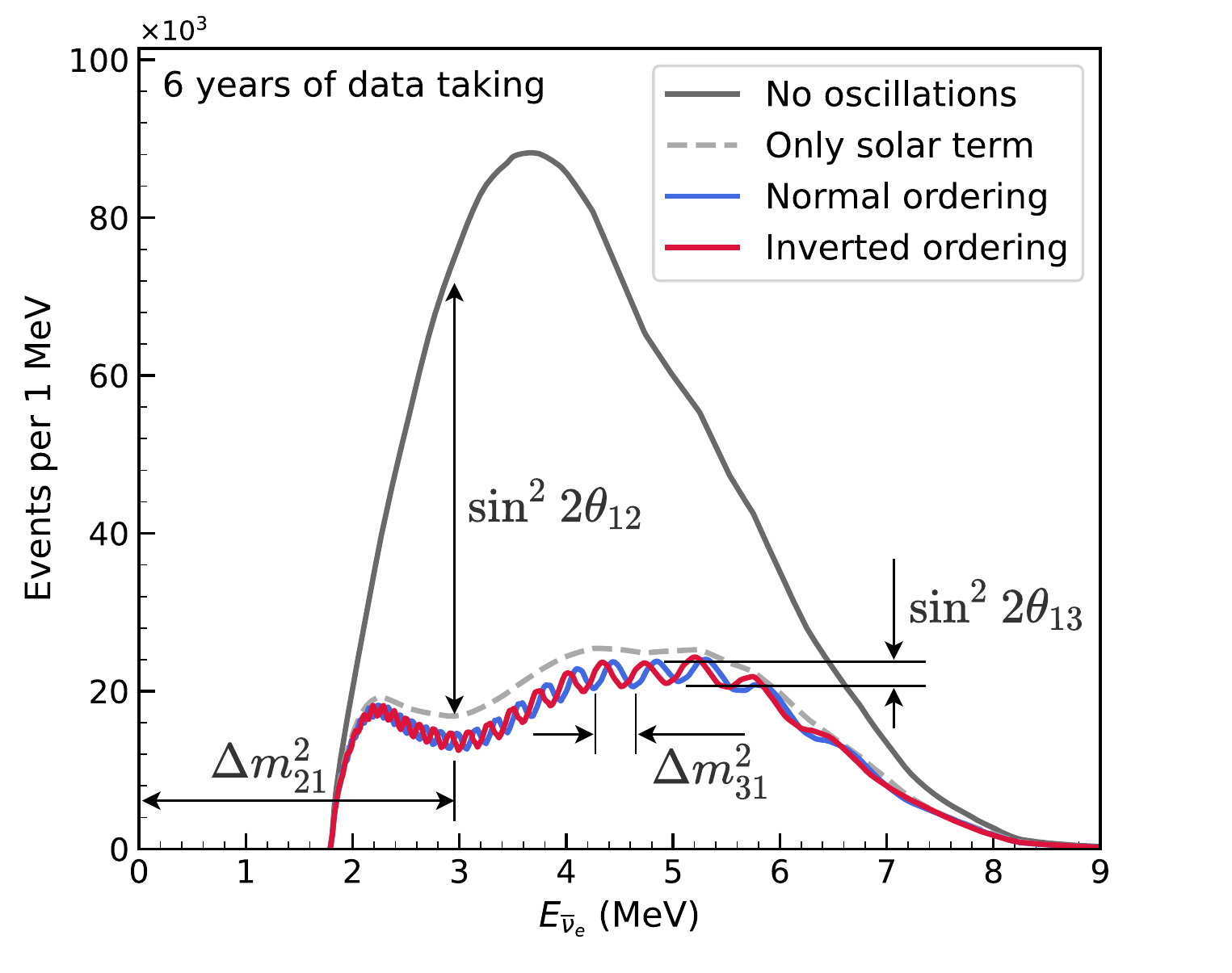}
  \caption{JUNO reactor antineutrino energy spectrum without (black) and with (grey, blue and red) the effect of neutrino oscillation. The reactor spectra are convoluted with the inverse beta decay cross-section and assume 6 years of data-taking. The gray dashed curve shows the spectrum when only the term in the disappearance probability that is modulated by $\sin^2 2\theta_{12}$ is included, whereas the blue and red curves are obtained using the full oscillation probability in vacuum for the normal and inverted mass orderings, respectively. A detector with perfect energy resolution is assumed for illustration purposes. Some spectral features driven by the $\Delta m^2_{31}$, $\Delta m^2_{21}$, $\sin^2 2\theta_{12}$, and $\sin^2 2\theta_{13}$ oscillation parameters are shown pictorially, illustrating the rich information available in a high-resolution measurement of the oscillated spectrum at JUNO's baseline. }
  \label{fig:spectrum}
\end{figure}

In JUNO's location, the energy spectrum will be distorted by a slow (low frequency) oscillation driven by \dmN\ and modulated by $\sin^2 2\theta_{12}$, as well as a fast (high frequency) oscillation driven by \Dm\ and modulated by $\sin^2 2\theta_{13}$, as shown in Fig.~\ref{fig:spectrum}. JUNO will be the first experiment to observe these two oscillation modes simultaneously. As detailed later, fitting the data spectrum against the predicted spectrum distorted by standard neutrino oscillations enables measuring the \Dm, \dmN, $\sin^2 \theta_{12}$, and $\sin^2 \theta_{13}$ oscillation parameters. The oscillated spectrum in JUNO also changes subtly depending on the neutrino mass ordering, thus providing sensitivity to this parameter. As previously mentioned, this measurement is not addressed in this publication.

\subsection{The JUNO Detector}
\label{sec:detector}

The JUNO detector will be deployed in an underground laboratory under the Dashi hill to limit the cosmogenic background. The 650 m overburden with average rock density of 2.61~g/cm$^3$ will suppress the cosmic-ray muon flux to $4.1\times10^{-3}$/(s$\cdot$m$^2$).

\begin{figure}[!h]
  \centering
  \includegraphics[width=0.9\textwidth]{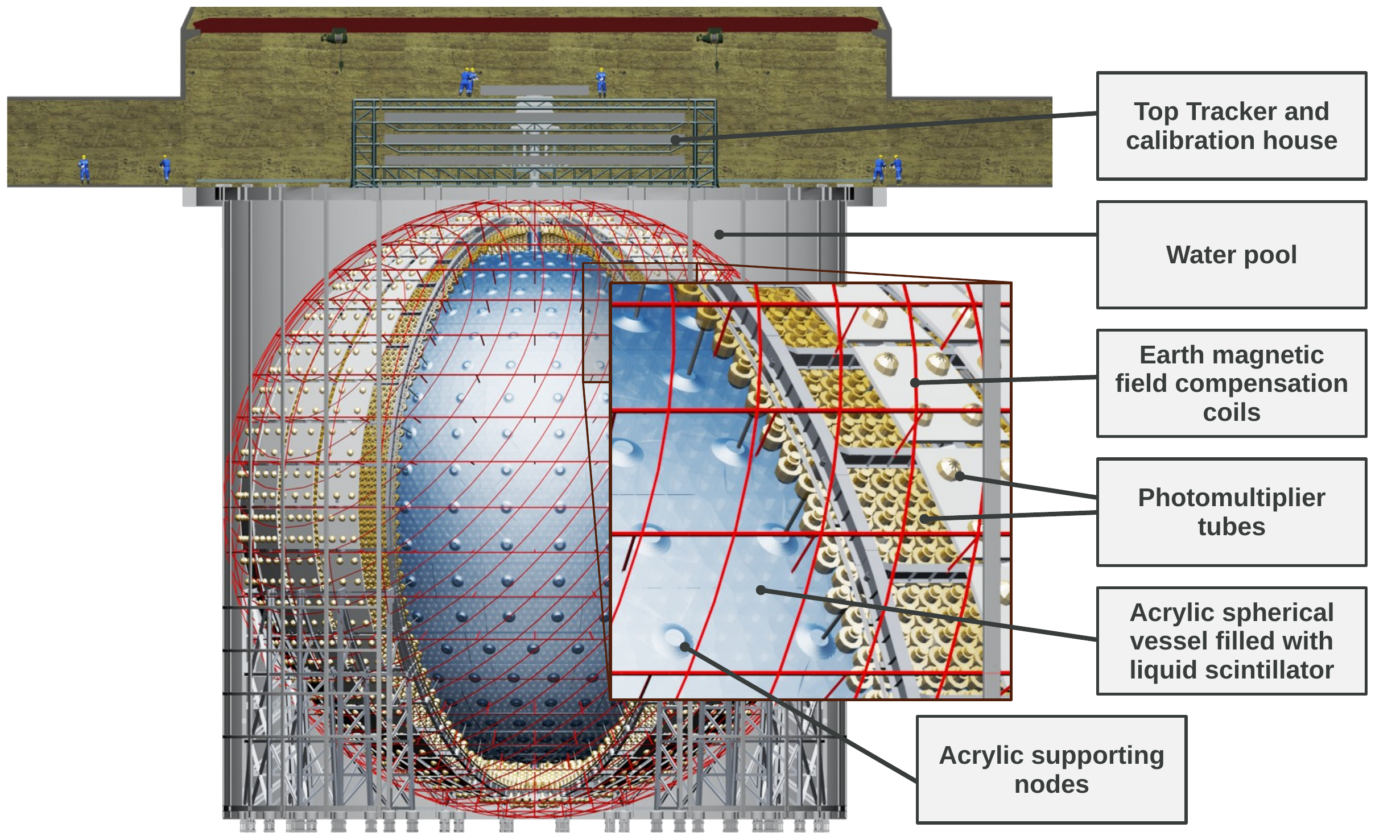}
  \caption{Schematic of the main JUNO detector. An acrylic sphere containing 20~kton of liquid scintillator is immersed in water and surrounded by 17,612 large (20-inch) and 25,600 small (3-inch) inward-facing PMTs. This Central Detector is optically decoupled from a surrounding water pool instrumented with 2,400 20-inch PMTs providing shielding and cosmic-ray muon tagging. A Top Tracker system consisting of three layers of plastic scintillator provides precision tracking of cosmic-ray muons entering the Central Detector. A calibration house is used to store the corresponding instruments deployed in the detector. Two sets of large coaxial coils running along different axes surround the Central Detector, largely suppressing the effects of the Earth's magnetic field on the 20-inch PMTs' collection efficiency.}
  \label{fig:detector}
\end{figure}

The main JUNO detector is shown in Fig.~\ref{fig:detector}. The primary antineutrino target of 20~kton of liquid scintillator is contained in a transparent 12-cm thick acrylic sphere 35.4~m in diameter. This constitutes the largest detector of this kind, securing JUNO's desired antineutrino statistics. Disentangling the two oscillation modes requires the detector to have the ability to measure the fast atmospheric oscillations, for which an unprecedented energy resolution is required.
The acrylic sphere is surrounded by 17,612 large 20-inch photomultiplier tubes (PMTs), referred to as LPMTs, and 25,600 small 3-inch PMTs, referred to
as SPMTs, yielding an integral 77.9\% photo-cathode coverage~\cite{JUNO:2022hxd}.
An ultra-pure water buffer with $\sim$1.5\,m thickness fills the volume between the acrylic and the LPMT photocathodes. The light yield at the detector center is expected to be $\sim$1345~photoelectrons (PEs) per MeV~\cite{JUNO:2020xtj}.
This represents at least 2.5$\times$ more light compared to the second highest yield achieved with the same technology~\cite{BOREXINO:2020aww}, and provides the required $\leq$3\,\% at 1~MeV energy resolution for the MO determination~\cite{Djurcic:2015vqa}.
The main detector is fully surrounded by an ultra-pure water Cherenkov detector that serves as both an active veto for cosmic muons (efficiency \textgreater99.5\%) and as a passive shield against external radioactivity and neutrons from cosmic rays. The minimal thickness of this detector is 2.5~m.
The muon cosmic veto system is supplemented with an external muon tracker consisting of three layers of plastic scintillator repurposed from the OPERA experiment~\cite{ADAM2007523} located at the top and providing a muon track angular reconstruction precision of 0.20\,$\degree$. This system covers about 60\% of the surface above the water pool. More details about JUNO's detector design can be found in Ref.~\cite{JUNO:2022hxd}. Discussion on the detector response, the corresponding systematic uncertainties, and their impact in this analysis, is deferred to Section~\ref{subsec:det_res}.

\subsection{The JUNO Nuclear Reactors}
\label{sec:JUNO_Reactor}
As shown in Fig.~\ref{fig:setup}, the primary reactor antineutrino sources for JUNO are the Taishan and Yangjiang NPPs, with two and six cores respectively, located at an average distance of 52.5 km. The next closest reactor complex to JUNO is Daya Bay, whose antineutrino flux slightly degrades the sensitivity to the oscillation parameters and is thus considered in the analysis. The reactor power, baselines, and expected IBD rates from Taishan, Yangjiang, and Daya Bay reactor cores, are summarized in Table~\ref{tab:Reactor}. The Huizhou NPP, at a distance of 265~km, is still under construction but will not be ready until several years after the start of data taking. Given the uncertainty on its schedule, it is not considered in the analysis. Other NPPs are more than 300~km away and contribute approximately one event per day to the total IBD rate in JUNO. As discussed in Section~\ref{subsec:sel_bkg}, they are treated as a background. More information on the reactor antineutrino flux prediction and the associated systematic uncertainties can be found in Section~\ref{sec:Reactor_nu_Flux}.

\begin{table}[h!]
\begin{center}
\begin{tabular}{lcccc}
    \hline
    Reactor   & Power (GW$_{\rm th}$) & Baseline (km) & IBD Rate (day$^{-1}$) & Relative Flux (\%) \\
    \hline
    Taishan & 9.2   & 52.71 & 15.1  & 32.1  \\
    ~~~Core 1 & 4.6   & 52.77 & 7.5  & 16.0  \\
    ~~~Core 2 & 4.6   & 52.64 & 7.6  & 16.1  \\
    Yangjiang & 17.4  & 52.46 & 29.0  & 61.5  \\
    ~~~Core 1 & 2.9   & 52.74 & 4.8  & 10.1  \\
    ~~~Core 2 & 2.9   & 52.82 & 4.7  & 10.1  \\
    ~~~Core 3 & 2.9   & 52.41 & 4.8  & 10.3  \\
    ~~~Core 4 & 2.9   & 52.49 & 4.8  & 10.2  \\
    ~~~Core 5 & 2.9   & 52.11 & 4.9  & 10.4  \\
    ~~~Core 6 & 2.9   & 52.19 & 4.9  & 10.4  \\
    Daya Bay & 17.4  & 215  & 3.0  & 6.4 \\ \hline
\end{tabular}
\caption{\small {Characteristics of NPPs and their reactor cores considered in this analysis: the two closest ones to JUNO, Taishan and Yangjiang, at an approximate distance of 52.5~km, and the next closest, Daya Bay. The IBD rates are estimated from the baselines, full thermal power of the reactors, selection efficiency, and current knowledge of the neutrino oscillation parameters. Relative contribution to the total antineutrino signal in JUNO is indicated in the last column.
}}
	\label{tab:Reactor}%
\end{center}
\end{table}%

\section{High Precision Reactor Antineutrino Detection}

\subsection{Reactor Antineutrino Selection and Residual Backgrounds}
\label{subsec:sel_bkg}
Reactor antineutrinos in JUNO are detected through the inverse beta decay (IBD) reaction
$
\bar{\nu}_e + p \to e^{+} + n
$.
The kinetic energy deposited by the positron via ionisation, together with its subsequent annihilation into typically two 
0.511\,MeV photons, forms a prompt signal. The impinging neutrino transfers most of its energy to the positron. This allows the deposited visible energy of the positron to be directly and very accurately related to the antineutrino energy, which is the relevant  metric for neutrino oscillation measurements. The neutron is captured in an average of $\sim$220~$\mu$s, and the corresponding photon emission forms a delayed signal. The neutron is captured dominantly on hydrogen~($\sim$99\%), releasing a single 2.2~MeV photon, and very infrequently on carbon~($\sim$1\%), yielding a gamma-ray signal with 4.9~MeV of total energy.
With a typical kinetic energy ranging from zero to a few tens of keV, the neutron in the IBD interaction carries only a small fraction of the initial antineutrino energy. However, due to the unprecedented energy resolution of JUNO, neutron recoils cannot be neglected, and the differential IBD cross-section is used in our calculations. We have adopted the IBD cross-section from~Ref.~\cite{Vogel:1999zy}. The small uncertainty in this quantity has no appreciable impact on the results presented in this publication.

The IBD prompt-delayed spatial and temporal coincidence signature can be mimicked by other events in the detector, giving rise to backgrounds. There are four main sources:
\begin{itemize}
    \item Radiogenic events, i.e. $\alpha$, $\beta$, $\gamma$ decays from natural radioactivity in the material of the detector.
    \item Cosmogenic events, i.e. fast neutrons and unstable isotopes produced by impinging muons on $^{12}$C, typically via spallation.
    \item Atmospheric neutrinos, i.e. neutrinos of all flavors created in the reactions set about by the collision of primary cosmic rays with the Earth's atmosphere.
    \item Electron antineutrinos emitted by distant reactors or created in the U and Th decay chains in Earth, i.e. geoneutrinos.

\end{itemize}

The coincidence of two otherwise uncorrelated events, typically of radiogenic origin, forms the so-called accidental background.
This background dominates the low energy part of the spectrum due to its nature. However, it can be measured with an excellent precision, typically at the permille level, and subtracted by off-time window techniques, as demonstrated by current-generation reactor antineutrino experiments~\cite{Adey:2018zwh,DoubleChooz:2019qbj,Bak:2018ydk}.

Correlated backgrounds are by definition produced by a single physics process and yield both a prompt and a delayed signal. The most important such backgrounds are cosmogenic $^9$Li/$^8$He and fast neutrons, which can only be further suppressed by increasing the overburden. There are also geoneutrinos, mostly below 2.5\,MeV in antineutrino energy~\cite{Sramek:2012hma}, and atmospheric neutrinos~\cite{Cheng:2020aaw}. The latter can produce neutrons, protons, $\alpha$ particles, and excited light nuclei that deposit their energy immediately or shortly after production and can thus mimic the IBD signature when followed by a neutron capture. There is only one radiogenic process leading to a correlated background deserving consideration: the ${}^{13}$C($\alpha$, $n$)${}^{16}$O reaction in the liquid scintillator. This background is expected to be small in JUNO, more so given the stringent radiopurity control that is envisaged~\cite{JUNO:2021kxb}. The production of fast neutrons and gamma rays via spontaneous fissions and $(\alpha, n)$ reactions in peripheral materials of the detector~\cite{PhysRevD.104.092006} is expected to have a negligible contribution to this analysis.

\begin{figure}[!h]
  \centering
  \includegraphics[width=0.9\textwidth]{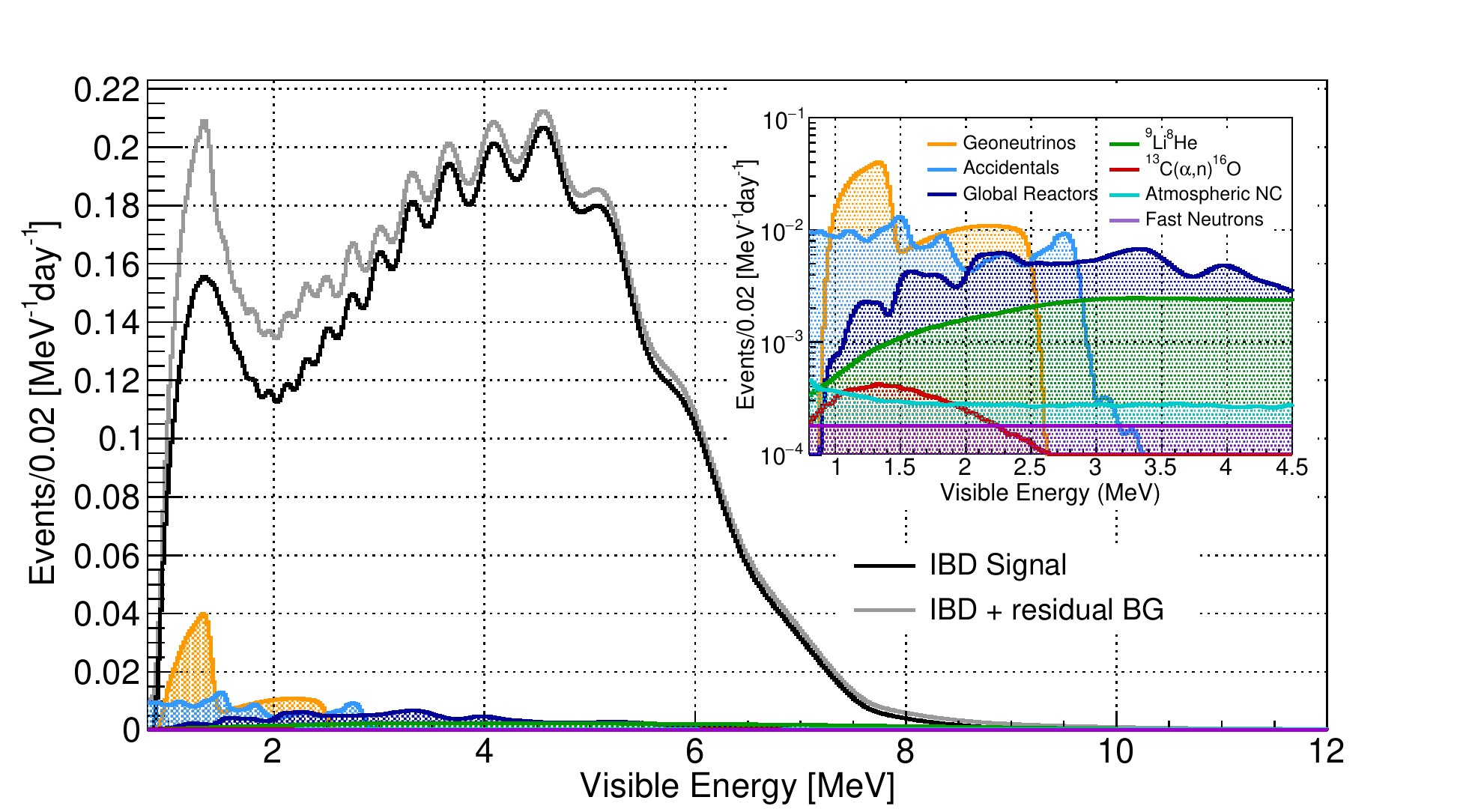}
  \caption{
  Visible energy spectrum expected in JUNO as measured with the LPMT system with (grey) and without (black) backgrounds. The assumptions detailed in the text are used, which include the energy resolution from Ref.~\cite{JUNO:2020xtj}. The inset shows the spectra of the expected backgrounds, which amount to about 7\% of the total IBD candidate sample and are mostly localized below $\sim$3~MeV.}
  \label{fig:IBD}
\end{figure}

IBD selection criteria are designed to suppress the aforementioned backgrounds while keeping a high efficiency for true reactor antineutrino IBD events. First, prompt and delayed candidate events are restricted to the energy windows
[0.7,~12.0]~MeV and [1.9,~2.5] $\cup$ [4.4,~5.5]~MeV, respectively. IBD events are expected to dominate the [0.7,~8.0]~MeV prompt energy range, as shown in Fig.~\ref{fig:IBD}.
The delayed signal energy selection windows are selected to be centered around 
2.2~MeV and 4.9~MeV
corresponding to neutron capture on hydrogen and carbon, respectively. Prompt or delayed events are discarded if their vertices are more than 17.2~m far away from the detector center, since the external background rate is larger at the edge of the acrylic sphere. This fiducial volume cut will be further optimised upon data taking based on the final radiopurity of the PMTs and the detector materials.
To further reduce the accidental background, the surviving prompt-delayed pairs are restricted to occur with a time separation $\Delta T_{p-d}$ smaller than 1.0~ms and a spatial 3D separation $\Delta R_{p-d}$ smaller than 1.5~m.

A series of cosmic muon veto cuts are enforced to suppress the cosmogenic backgrounds, most of which satisfy the IBD coincidence selection criteria. Muon-induced neutrons can be greatly reduced by imposing a time cut proportional to the characteristic time of neutron capture, as done in other underground liquid scintillator experiments~\cite{Adey:2018zwh,DoubleChooz:2019qbj,Bak:2018ydk,KamLAND:2008dgz}. However, this approach does not fully eliminate the longer-lived isotopes, in particular $^9$Li/$^8$He, that are produced along the muon track. The exploitation of this topological correlation has been considered by other experiments~\cite{KamLAND:2008dgz,Super-Kamiokande:2016yck}.
A refinement of this strategy has been developed for JUNO with state of the art simulations by using a different veto time window depending on the candidate event's proximity to a recent muon track or spallation neutron capture. This strategy is a variation of the muon veto reported in Ref.~\cite{Abusleme:2020zyc}, but optimized for the IBD selection. The details are as follows:
\begin{itemize}
    \item
    For all muons passing the water pool Cherenkov detector and/or the Central Detector, a veto of 1~ms after each muon is applied over the whole fiducial volume to suppress spallation neutrons and short-lived radio-isotopes.

    \item
    For well-reconstructed muon tracks in the Central Detector caused by single or two far-apart muons, a veto of 0.6~s, 0.4~s, and 0.1~s is applied to candidate events with reconstructed vertices smaller than 1~m, 2~m, and 4~m away from the track(s), respectively.

    \item For events containing two close and parallel muons (\textless 3~m), which constitute roughly 0.6\% of muon-related events, a single track is often reconstructed. The veto is applied around this track as described above, but the cylinder radii are increased according to their separation, which can be inferred from the charge pattern around the entrance and exit points.

    \item
    For events where a track cannot be properly reconstructed, which amount to about 2\% of all muon-related events and occur primarily when more than two muons go through the detector simultaneously, a 0.5~s veto is applied over the whole fiducial volume.

    \item
    A 1.2~s veto is applied on any candidate events reconstructed inside a 3~m radius sphere around spallation neutron capture events. This cut helps further reject backgrounds from cosmogenic isotope decays.

\end{itemize}
This updated strategy improves the muon veto efficiency from the previously reported 83.0\%~\cite{An:2015jdp} to 91.6\%, while keeping the residual cosmogenic background almost unchanged. The combined antineutrino detection efficiency after all selection cuts is 82.2\%. A rounded value of 82.0\% was used in the analyses reported here. Breakdown of the selection efficiency is summarised in Table~\ref{tab:efficiency}.

\begin{table}[!h]
    \centering
    \begin{tabular}{lccc}
    \hline
    Selection Criterion & Efficiency (\%) & IBD Rate (day$^{-1}$)\\
    \hline
    All IBDs						& 100.0 	& 57.4 \\
    Fiducial Volume 						& 91.5 	& 52.5 \\
    IBD Selection 				& 98.1		& 51.5 \\
    ~~~~Energy Range 					& ~~~~99.8		& - \\
    ~~~~Time Correlation ($\Delta T_{p-d}$) 		& ~~~~99.0 	& - \\
    ~~~~Spatial Correlation ($\Delta R_{p-d}$) 	& ~~~~99.2		& - \\
    Muon Veto (Temporal$\oplus$Spatial)		& 91.6 	& 47.1  \\
    \hline
    Combined Selection					& 82.2 	&  47.1\\
    \hline
    \end{tabular}
    \caption{Summary of cumulative reactor antineutrino selection efficiencies. The
    reported IBD rates refer to the expected events per day after the selection criteria are progressively applied. These rates are calculated for nominal reactor power, and do not include any reactor time off.  
    }
    \label{tab:efficiency}
\end{table}

After applying the antineutrino event selection cuts mentioned above, seven backgrounds remain that are considered in this analysis: geoneutrinos, $\bar\nu_e$'s from world reactors (with a baseline to JUNO larger than 300~km), accidental coincidences, $^9$Li/$^8$He decays, atmospheric neutrinos, fast neutrons, and ${}^{13}$C($\alpha$, $n$)${}^{16}$O interactions.
Their rates and uncertainties are summarized in Table~\ref{tab:background}. These values are consistent with those in our previous work~\cite{An:2015jdp}, although some adjustments are made. The rates of geoneutrinos and $^9$Li/$^8$He decays are adjusted by +0.1~day$^{-1}$ and -0.8~day$^{-1}$, respectively, because of the new muon veto strategy. Likewise, the accidental background rate is reduced by 0.1~day$^{-1}$ due to new knowledge on the radiopurity of the detector components~\cite{JUNO:2021kxb}. The world reactors and the atmospheric neutrino backgrounds are new additions in this publication. The former is calculated from Ref.~\cite{Dye:2015bsw} and the same uncertainty of the $\bar{\nu}_e$ signal described in Section~\ref{sec:Reactor_nu_Flux} is applied. The latter is estimated following the methodology of Ref.~\cite{Cheng:2020aaw}. The IBD selection criteria is applied to simulate final states of atmospheric neutrinos interacting with $^{12}$C nuclei in the liquid scintillator. In the [0.7,~12.0]~MeV energy range, neutral-current interactions are found to dominate, with charged-current interactions contributing a negligible amount. The uncertainty is estimated from the largest variation in predicted rate between an interaction model that relies on GENIE 2.12.0, which is taken as the nominal, and four others relying on the NuWro generator that use different nuclear models and values of the axial mass~\cite{Cheng:2020aaw}.

The geoneutrino and world reactors' antineutrino spectra are obtained from Refs.~\cite{Araki:2005qa} and~\cite{Dye:2015bsw}, respectively. The accidental spectrum is obtained by applying the IBD selection to events from a full JUNO simulation with a recently re-estimated radioactivity budget~\cite{JUNO:2021kxb}. The $^9$Li/$^8$He spectrum is obtained from a theoretical calculation. The atmospheric neutrino spectrum is the one produced by the nominal interaction model relying on the GENIE 2.12.0 generator in Ref.~\cite{Cheng:2020aaw}. The fast neutron spectrum is assumed to be flat in the energy range of interest, which is a reasonable approximation as seen in both simulation and recent reactor experiments~\cite{DoubleChooz:2019qbj,Bak:2018ydk,DayaBay:2016ggj}. Finally, the spectrum of ${}^{13}$C($\alpha$,n)${}^{16}$O is obtained from simulation~\cite{DayaBay:2016ggj}. In all cases the full detector response of Section~\ref{subsec:det_res} is applied.

With the exception of the two newly considered backgrounds, the spectral shape uncertainties are the same as in Ref.~\cite{An:2015jdp}. The shape uncertainty of the world reactors' $\bar{\nu}_e$ background is considered to be the same of the $\bar{\nu}_e$ signal, described in Section~\ref{sec:Reactor_nu_Flux}. The spread between interaction models is assigned as the shape uncertainty of the atmospheric neutrino background. All shape uncertainties are conservatively treated as bin-to-bin uncorrelated. Their knowledge is expected to improve through JUNO data analysis.

Compared to other underground liquid scintillator experiments, the impact of the backgrounds on the precision of the measurement of the oscillation parameters is limited. This is because JUNO exploits the large spectral shape distortion of the IBD spectrum as the primary handle to extract the oscillation parameters. As illustrated in Fig.~\ref{fig:IBD}, the residual backgrounds' spectra are manifestly distinct from the oscillated spectrum.

\begin{table}[!t]
    \centering
    \begin{tabular}{cccc}
    \hline
    Background & Rate (day$^{-1}$) & Rate Uncertainty (\%) & Shape Uncertainty (\%)\\
    \hline
    Geoneutrinos 		& 1.2 & 30 & 5 \\
    World reactors 		& 1.0 & 2 & 5 \\
    Accidentals 		& 0.8 & 1 & negligible \\
    $^9$Li/$^8$He 	    & 0.8 & 20 & 10 \\
    Atmospheric neutrinos 		& 0.16 & 50 & 50 \\
    Fast neutrons 		& 0.1 & 100 & 20 \\
    ${}^{13}$C($\alpha$,n)${}^{16}$O & 0.05 & 50 & 50 \\
    \hline
    \end{tabular}%
    \caption{Background rates and uncertainties.}
    \label{tab:background}%
\end{table}%

\subsection{Detector Response}
\label{subsec:det_res}
The extraction of the oscillation parameters relies strongly on the careful control of systematic uncertainties affecting both the precision and accuracy of the spectral distortion caused by neutrino oscillation. The energy response model considered in this analysis includes three effects: energy transfer in the IBD reaction, detector nonlinearity, and energy resolution. The event-vertex dependence of the energy response, i.e. the non-uniformity, also plays an important role and has been included in the energy resolution model as described below.

Energy transfer in the IBD reaction is calculated by integrating the IBD differential cross-section over the positron scattering angle. The kinetic energy of the positron, together with the typically two 0.511~MeV annihilation photons, is assumed to be fully deposited in the detector and is defined as $E_{\rm dep}$. Energy losses from escaping secondary particles only affect less than 1\% of IBD events and are consequently ignored in this analysis.
 Due to the quenching effect of the scintillation light, the Cherenkov radiation, and the photon detection, the visible energy that would be observed if JUNO had perfect energy resolution, defined as $E^*_{\rm vis}$, does not depend linearly on the deposited energy. Accordingly, the factor $E^*_{\rm vis}/E_{\rm dep}$ represents the nonlinear response of the detector. The instrumental charge nonlinearity of the JUNO PMTs and their electronics is assumed to be negligible (\textless0.3\%) thanks to the dual-calorimetry calibration technique described in Ref.~\cite{JUNO:2020xtj}. Therefore, in this analysis only the nonlinearity from the liquid scintillator itself is considered and assumed to be identical to the one measured in the Daya Bay experiment~\cite{Adey:2019zfo:DYB_NL}, whose scintillator composition is similar. The implementation of this systematic uncertainty in JUNO follows a similar strategy as in Daya Bay, where a nominal curve is first employed and four curves weighted by pull parameters are used to account for possible variations and to generate an uncertainty band, as shown in Ref.~\cite{Adey:2019zfo:DYB_NL}.

Finally, the visible energy $E_{\rm vis}$ will be further smeared relative to $E^*_{\rm vis}$ because of the finite energy resolution of the detector. When detector leakage effects are neglected, the resolution can be parameterized using a Gaussian function with a standard deviation $\sigma_{E_{\rm vis}}$ given by
\begin{equation}
	\frac{\sigma_{E_{\rm vis}}}{E_{\rm vis}}=\sqrt{\left(\frac{a}{\sqrt{E_{\rm vis}}}\right)^2+b^2+\left(\frac{c}{E_{\rm vis}}\right)^2}\,,
    \label{eq:EnergyResolution:abc}
\end{equation}
where $a$ is the term driven by the Poisson statistics of the total number of detected photoelectrons, $c$ is dominated by the PMT dark noise, and $b$ is dominated by the detector's spatial non-uniformity. The large and small PMTs work as two complementary photon detection systems, resulting in the energy of every event being measured twice with very different resolutions. The energy resolution of the LPMT system was carefully studied using Monte Carlo simulation in Ref.~\cite{JUNO:2020xtj}, yielding $a=(2.61\pm0.02)\%\sqrt{\textrm{MeV}},\:b=(0.82\pm0.01)\%,$ and $c=(1.23\pm0.04)\%\: \textrm{MeV}$ as the average values for the fiducial volume of the detector. For the SPMT system, $a$ is expected to dominate because of the smaller light level, making $b$ and $c$ almost irrelevant. The value of this parameter is determined according to the ratio of the simulated total number of photoelectrons between the two PMT systems, while $c$ is calculated based on the measured dark noise rate of the SPMTs~\cite{Cao:2021wrq}. $b$ is not modified because detector effects are expected to be largely the same for both PMT systems. This results in $a=15.36\%\sqrt{\textrm{MeV}},\:b=0.82\%,$ and $c=6.77\%\: \textrm{MeV}$ for the SPMT energy resolution. Despite the poorer energy resolution, the SPMT system allows for a semi-independent measurement of the slow $\Delta m^2_{21}$-$\sin^2 \theta_{12}$ oscillation, as explained in Section~\ref{subsec:results}.

\begin{figure}[!h]
  \centering
  \includegraphics[width=0.9\textwidth]{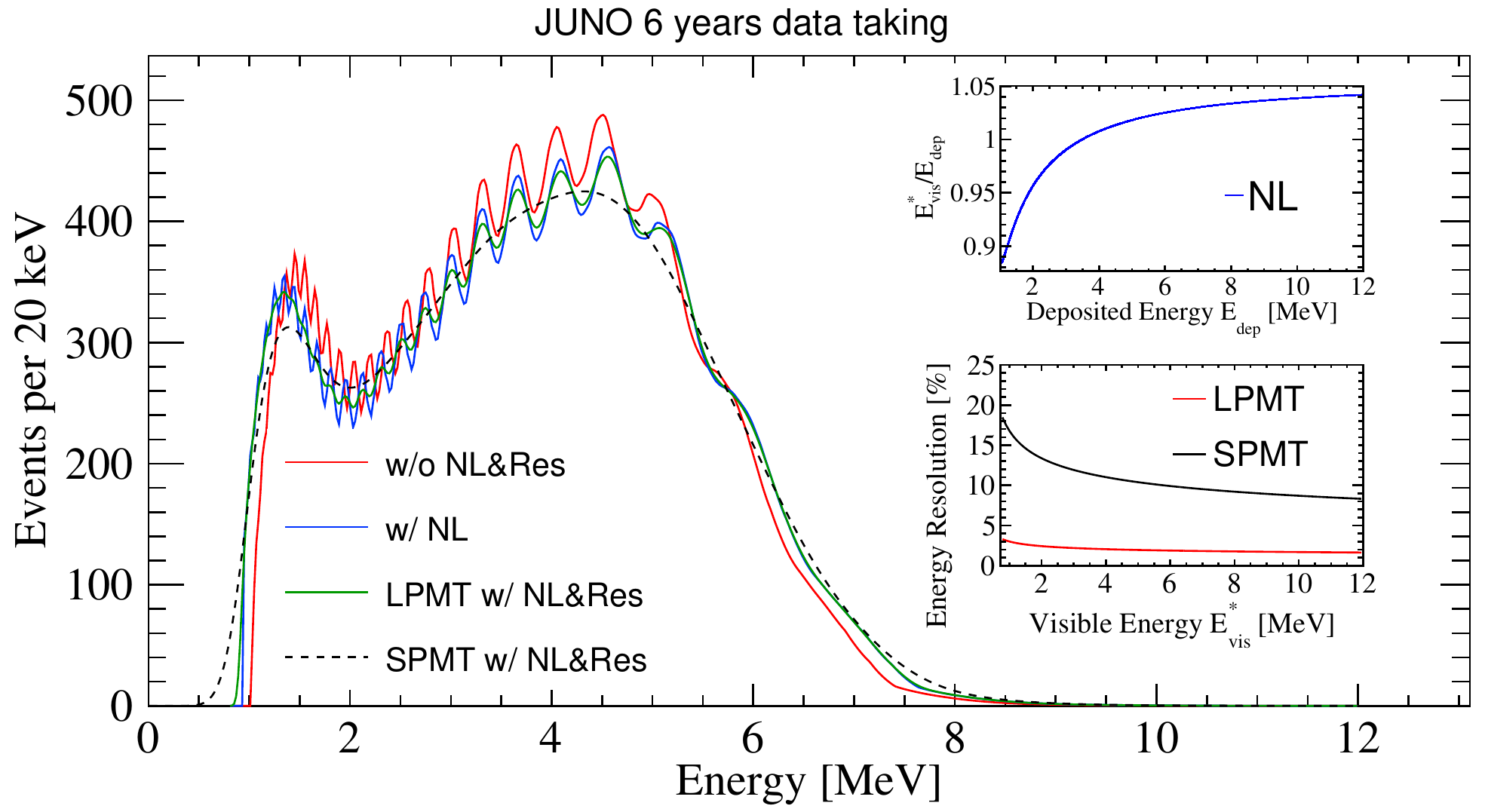}
  \caption{Detector response impact on the IBD spectrum. The top right panel shows the nonlinear energy response of the liquid scintillator. The bottom right panel shows the energy resolution of the LPMT and the SPMT systems as a function of visible energy. In both cases the resolution is described using the same model introduced in Eq.~\eqref{eq:EnergyResolution:abc}, with $a$=2.61\%, $b$=0.82\%, $c$=1.23\% for LPMT and $a$=15.36\%, $b$=0.82\%, $c$=6.77\% for SPMT. The main panel shows the deposited energy spectrum from the IBD reaction in 6~years of JUNO data without detector nonlinearity (NL) nor energy resolution (Res) in red, with NL only in blue, and with both detector effects in green, where the energy resolution corresponds to the LPMT system. The spectrum detected by the SPMT system with both NL and Res is also shown in dashed black.}
  \label{fig:response}
\end{figure}

The energy spectrum at different stages in the calculation can be found in Fig.~\ref{fig:response}, embedded with the nonlinearity curve and the energy resolution curves of both PMT systems.

\subsection{Reactor Antineutrino Flux}
\label{sec:Reactor_nu_Flux}

The expected visible energy spectrum observed at JUNO can be calculated as
\begin{equation}
  \label{eq:energy_response}
\begin{split}
    S(E_{\rm vis}) = N_p\cdot\epsilon\cdot\int_{T_{\rm DAQ}}dt
    \int_{1.8~\mathrm{MeV}}^{12~\mathrm{MeV}}dE_{\bar{\nu}_e}
     \cdot \Phi(E_{\bar{\nu}_e},t)\cdot \sigma(E_{\bar{\nu}_e})
    \cdot R(E_{\bar\nu_e}, E_{\rm vis}),
\end{split}
\end{equation}
where $R(E_{\bar\nu_e},E_{\rm vis})$ is the detector energy response function embedding the effects described in Sections~\ref{subsec:det_res} that maps the antineutrino energy to the visible energy,
$\sigma(E_{\bar{\nu}_e})$ is the IBD cross-section,
$\Phi(E_{\bar{\nu}_e},t)$ is the oscillated reactor antineutrino flux in JUNO at time $t$, $T_{\rm DAQ}$ is the total data taking time, $\epsilon$ is the detection efficiency, and $N_p=1.44\times10^{33}$ is the number of free protons in the detector target.

In a commercial reactor, electron antineutrinos are produced from the fission products of four major isotopes, $^{235}$U, $^{238}$U, $^{239}$Pu, and $^{241}$Pu.
The oscillated antineutrino flux at time $t$ is written as
\begin{equation}\label{equ_generic}
    \Phi(E_{\bar{\nu}_e}, t)  = \sum_r \frac{{\cal P}_{\bar{\nu}_{e} \to \bar{\nu}^{}_e}(E_{\bar{\nu}_e},L_{r})}{ 4\pi L^2_{r}}\phi_r(E_{\bar{\nu}_e},t),
\end{equation}
where $E_{\bar{\nu}_e}$ is the $\bar\nu_e$ energy, $r$ is the reactor index,
$L_{r}$ is the distance from the detector to reactor $r$, ${\cal P}_{\bar{\nu}_{e} \to \bar{\nu}^{}_e}(E_{\bar{\nu}_e},L_{r})$ is the $\bar{\nu}_e$ survival probability,
and $\phi_r(E_{\bar{\nu}_e},t)$ is the reactor antineutrino energy spectrum. The latter can be calculated as
\begin{equation}\label{equ_prediction}
\phi_r(E_{\bar{\nu}_e},t) = \frac{W_r(t)}{\sum_i f_{ir}(t) e_i}\sum_i f_{ir}(t)s_i(E_{\bar{\nu}_e}),
\end{equation}
where $W_r(t)$ is the thermal power, $e_i$ is the mean energy released per fission for isotope $i$, $f_{ir}(t)$ is the fission fraction, and $s_i(E_{\bar{\nu}_e})$ is the antineutrino energy spectrum per fission for each isotope. Averaged reactor power and fission fractions are used for this study, although these quantities will be provided by the power plants for each core as a function of time once JUNO begins operating.
To account for refueling, which typically takes one month per year, the average reactor thermal power is calculated as the nominal value reduced by a reactor duty cycle factor of 11/12. The average fission fractions are assumed to be 0.58, 0.07, 0.30, 0.05, with mean energies per fission of 202.36~MeV, 205.99~MeV, 211.12~MeV, 214.26~MeV~\cite{Ma:2012bm} for $^{235}$U, $^{238}$U, $^{239}$Pu, $^{241}$Pu, respectively. The $\bar\nu_e$ energy spectrum per fission of $^{235}$U, $^{239}$Pu and $^{241}$Pu is obtained from Huber~\cite{Huber:2011wv}, and of $^{238}$U from Mueller \textit{et al.}~\cite{Mueller:2011nm}.

Additional corrections are applied to account for the non-equilibrium and spent nuclear fuel contributions.
The former arises from beta decays of some long-lived fission fragments and adds an extra $\sim$0.6\% to the antineutrino flux.
The latter is caused by the spent nuclear fuel removed to cooling pools near the reactor cores still emitting antineutrinos and contributes an additional $\sim$0.3\% to the flux.
The corrections are obtained from Ref.~\cite{An:2016srz},
both of which are assigned a 30\% rate uncertainty and a negligible spectrum shape uncertainty, in agreement with the latest results from Daya Bay~\cite{Adey:2018zwh}. The total unoscillated spectrum for JUNO is obtained by aggregating the contributions of the four isotopes in the Huber-Mueller model and correcting for these two effects. Discrepancies have been found between the data and the models, most notably a $\sim$5\% deficit of the total flux with respect to the Huber-Mueller prediction, commonly known as the reactor antineutrino anomaly, and a spectral distortion in the $\sim$[4,~6]~MeV region observed when comparing to both conversion and summation models~\cite{ DoubleChooz:2019qbj,Bak:2018ydk, DayaBay:2019yxq,Ko:2016owz,PROSPECT:2020sxr,STEREO:2020fvd,Giunti:2021kab,Hayes:2016qnu}. Therefore, the ratio between the measurement and the total prediction in Daya Bay~\cite{An:2016srz} is used to further correct the total prediction used in this sensitivity study.

The uncertainties of the predicted reactor antineutrino flux are listed at the top of Table~\ref{tab:ratesyst}.
The baselines are known to 1~m, resulting in a negligible contribution to the flux uncertainty at distances of $\sim$52.5~km.
The reactor power data will be provided by the NPPs with an uncertainty of 0.5\%. Likewise, the fission fractions will be provided with an uncertainty of 5\%, which will contribute an uncertainty of 0.6\% to the predicted number of events.
The mean energy per fission is known precisely and contributes only a 0.2\% uncertainty to the predicted number of events.
Finally, a 2\% correlated uncertainty is assigned for the mean cross-section per fission, which is the product of the IBD cross-section with the total antineutrino spectrum and is thus proportional to the number of predicted events.
All of these uncertainties are drawn directly from the experience accumulated by the Daya Bay experiment~\cite{An:2016srz}.

As noted in Section~\ref{sec:detector}, TAO is a satellite detector whose primary objective is to provide a precise and model-independent antineutrino spectrum for JUNO to use as a reference~\cite{Abusleme:2020bzt:TAO_CDR}.
This spectrum will be measured with sub-percent energy resolution in most of the energy region of interest.
The expected uncertainty from TAO's measured spectrum, estimated from a simulation of that detector~\cite{Abusleme:2020bzt:TAO_CDR}, is propagated as the spectral shape uncertainty in this analysis.
TAO will collect about two million IBD events in three years, representing 20 times the statistics of JUNO. As a result, the statistical uncertainty with 20 keV-sized bins will be below 1\% across the $\sim$[2.5,~5.5]~MeV energy range. The systematic uncertainties considered include the scintillator nonlinearity, differences in fission fractions, and the impact of using a fiducial volume cut. As will be shown in Fig.~\ref{fig:shape}, the combined statistical and systematic uncertainties yield a spectral shape uncertainty that is below $1.5\%$ in the [2,~4]~MeV energy region. This energy-dependent uncertainty replaces the uniform 1\% bin-to-bin uncertainty used in Ref.~\cite{An:2015jdp}.

\section{Neutrino Oscillation Analysis}

To extract the neutrino oscillation parameters, we compare the nominal spectrum, a proxy of the expected spectrum that JUNO will measure, illustrated in Fig.~\ref{fig:IBD}, against the hypothesis model based on the standard parametrization ($\Delta m^2_{31}$, $\Delta m^2_{21}$, $\sin^2 \theta_{12}$, and $\sin^2 \theta_{13}$) described in Section~\ref{sec:osc_framework}. The current section describes the procedure, inputs, and systematic uncertainties used to perform this comparison, as well as the resulting sensitivities. The sensitivities' evolution with time, the correlations between oscillation parameters, and the impact of the systematic uncertainties on the parameters' precision, are also shown.

\subsection{Statistical Method}
\label{sec:AnalysisFramwork:Stat}

To compare the data to the hypothesis model, we employ the least-squares method, and construct a binned $\chi^2$ with covariance matrices and/or pull terms to account for systematic uncertainties~\cite{Stump:2001gu},

\begin{equation}
	\chi^2\equiv\left(\boldsymbol{M-T(\theta,\alpha)}\right)^T\cdot \boldsymbol V^{-1}\cdot \left(\boldsymbol{M-T(\theta,\alpha)}\right)+\sum_{i}\left(\frac{\alpha_{i}}{\sigma_{i}}\right)^{2},
	\label{eq:chi_sq}
\end{equation}
where $\boldsymbol M$ and $\boldsymbol T$ represent the measured and expected vectors of events per individual energy bin, respectively, and $\boldsymbol V$ is the covariance matrix of the prediction. For this analysis, $\boldsymbol M$ is set to the nominal expectation without any fluctuations. $\boldsymbol T$ depends on the oscillation parameters $\boldsymbol \theta$ described in Section~\ref{sec:osc_framework}, as well as on the nuisance parameters $\alpha_{i}$, each of which has a corresponding systematic uncertainty $\sigma_i$. The pull terms on the right hand side of Eq.~\ref{eq:chi_sq} can substitute any covariance matrix representing a systematic uncertainty, and vice-versa.

The full analysis, from the determination of $\boldsymbol M$ and $\boldsymbol T(\theta,\alpha)$ to the sensitivity calculations, was independently carried out by four analysis groups that started from the same common inputs. These common inputs express the current best knowledge of JUNO's performance and reactor situation, as described in the previous sections, and are also used by other sensitivity studies within the JUNO collaboration. Each of the four groups chose a different strategy to perform the minimization of Eq.~\eqref{eq:chi_sq}. One group used a covariance matrix-only approach, two groups used a pull term-only approach, and a fourth group used a mixture of both. Results were carefully compared at every stage of the analysis chain and differences in the final sensitivities were found to be much smaller than the systematic uncertainties. Accordingly, only one set of results, which is representative of the four groups, is shown in the remainder of this publication.

\subsection{Rate and Shape Systematic Effects}
\label{sec:rate_shape_syst}

The assessment of the systematic uncertainties benefits largely from the large pool of knowledge accumulated by past and current reactor experiments, particularly those focused on precisely measuring the $\theta_{13}$ mixing angle~\cite{Adey:2018zwh,DoubleChooz:2019qbj,Bak:2018ydk}. Systematic effects fall into two categories: rate and shape. Rate systematic uncertainties are those affecting the total number of IBD candidates (normalization), while shape systematic uncertainties are those that can bias the expected spectral shape (events per individual energy bin).

\begin{table}[!h]
    \centering
    \begin{tabular}{lc}
	Component & Input Uncertainty (\%)\\
	\hline
	Flux 						& 2.2		\\
	~~~Baseline (L) 						& ~~~~~~-		\\
	~~~Energy per Fission  					& ~~~~~~0.2	\\
	~~~Thermal Power (P) 					& ~~~~~~0.5	\\
	~~~Fission Fraction 					& ~~~~~~0.6	\\
	~~~Mean Cross-Section per Fission		& ~~~~~~2.0	\\
	Detection 					& 1.0	\\
	~~~Fiducial volume (2~cm vertex bias)	& ~~~~~~0.4	\\
	~~~IBD Selection cuts        			& ~~~~~~0.2	\\
	~~~Muon Veto         	 				& ~~~~~~- 	\\
	~~~Proton Number 					& ~~~~~~0.9	\\
	Backgrounds 		& 1.0	\\
    	~~~Geoneutrinos   					& ~~~~~~0.8	\\
    	~~~$^9$Li/$^8$He 					& ~~~~~~0.4	\\
    	~~~Atmospheric neutrinos 	& ~~~~~~0.2	\\
    	~~~Fast neutrons 					& ~~~~~~0.2	\\
    	~~~${}^{13}$C($\alpha$,n)${}^{16}$O & ~~~~~~0.1	\\
    	~~~Accidentals 						& ~~~~<0.1	\\
    	~~~World reactors    	& ~~~~<0.1	\\
	\hline
    \end{tabular}
    \caption{Signal normalization systematic uncertainties of JUNO. All uncertainties (backgrounds included) are relative to the signal rate of 43.2 measured IBDs per day, which accounts for the reactors' duty cycle. These uncertainties are used as inputs to the analysis. The flux systematic uncertainties have correlated and uncorrelated terms, with respect to the reactors. See the text for more details. The detection systematic uncertainties contain the same items of Table~\ref{tab:efficiency} plus the uncertainty on the number of target protons.
    }
    \label{tab:ratesyst}%
\end{table}

Rate systematic effects and their corresponding uncertainties are summarized in Table~\ref{tab:ratesyst}. They are divided into three main subcategories: flux, detection, and backgrounds. Within the flux subcategory, reactor-related uncertainties impact the analysis differently depending on whether they are correlated (2\%) or uncorrelated (0.8\%). Reactor correlated uncertainties, namely the mean cross-section per fission and energy per fission, affect all reactors contributing to the total neutrino flux in the same way, while reactor uncorrelated uncertainties, namely the thermal power and fission fraction, can vary independently from reactor to reactor. The detection systematic uncertainty of 1\% encapsulates those uncertainties affecting the total number of selected IBD events.
The dominant contribution on the flux category is the mean cross-section per fission, while on the detection category it is the target proton number uncertainty, which is estimated based on Daya Bay's experience~\cite{DayaBay:2016ggj}. The background rate uncertainties that are used as input to the analysis are shown in Table~\ref{tab:background}, but
Table~\ref{tab:ratesyst} shows the relative uncertainty of the background rates compared to the IBD signal rate so they can be compared to other rate systematic uncertainties. The relative impact of the various backgrounds is different in JUNO compared to short baseline reactor neutrino experiments because of the drastic difference in the signal to background ratio. The two dominant backgrounds in terms of their uncertainty relative to the IBD signal are geoneutrinos and $^{9}$Li$/^{8}$He, but their very different spectral shapes compared to the distorted IBD spectra, illustrated in Fig.~\ref{fig:IBD}, provide additional constraints during the analysis.

\begin{figure}[!h]
  \centering
    \includegraphics[width=0.9\textwidth]{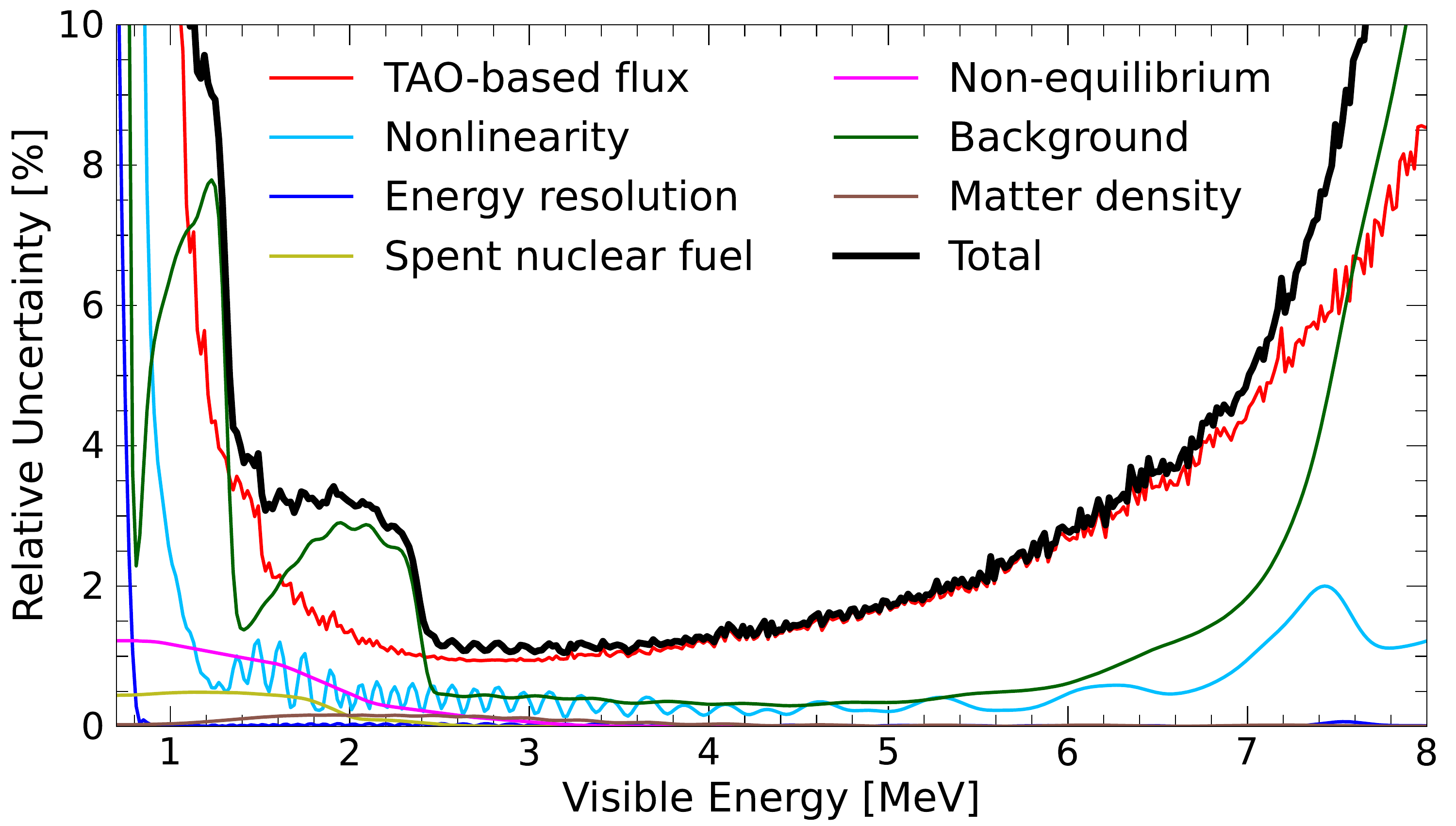}
  \caption{Shape uncertainties relative to the number of events in each bin. These are obtained by generating simulated samples where systematic parameters are varied based on their assumed uncertainties and taking the ratio of the diagonal elements of the resulting covariance matrix to the number of nominal reactor antineutrino signal events in each bin. The rate uncertainties of the spent nuclear fuel and non-equilibrium corrections, as well as of the backgrounds, also distort the observed spectrum, and are consequently included in this figure.}
  \label{fig:shape}
\end{figure}

The effects distorting the shape of the spectrum and their impact relative to the number of events are summarized in Fig.~\ref{fig:shape}.
The main contributions are the  uncertainties in the reactor antineutrino spectrum and the backgrounds. As already mentioned in Section~\ref{sec:Reactor_nu_Flux}, there is mounting evidence that the systematic uncertainties of the predicted reactor antineutrino flux and shape are underestimated.
For JUNO, the dedicated satellite detector TAO will provide the data-driven spectral uncertainty with an unprecedented energy resolution better than 2\% at 1~MeV~\cite{Abusleme:2020bzt:TAO_CDR}. We use this TAO-based spectrum model uncertainty in our analysis. 
The uncertainty of the detector response model, described in Section~\ref{subsec:det_res} and typically controlled to less than 0.5\%~\cite{Adey:2019zfo:DYB_NL,Chauveau:2018ze,Re:2018ze,Gando:2018ze}, is also important for the accuracy of the neutrino oscillation parameters. Its propagated uncertainty relates directly to the signal spectral shape, thus the small oscillations on Fig.~\ref{fig:shape}.
This figure also shows the background uncertainty with respect to the antineutrino signal, which includes all rate and shape uncertainties of Table~\ref{tab:background}. Similarly, the non-equilibrium and spent nuclear fuel uncertainties, discussed in the end of Section~\ref{sec:Reactor_nu_Flux}, are included in this figure, since they affect the signal spectrum in specific energy ranges. Finally, the 6\% uncertainty on the matter density impacts the oscillation probability, as described in Section~\ref{sec:NO_Matter}, but makes a very small contribution to the shape uncertainty.

\subsection{Neutrino Oscillation Sensitivity Results}
\label{subsec:results}

\begin{figure}[!h]
  \centering
    \includegraphics[width=0.9\textwidth]{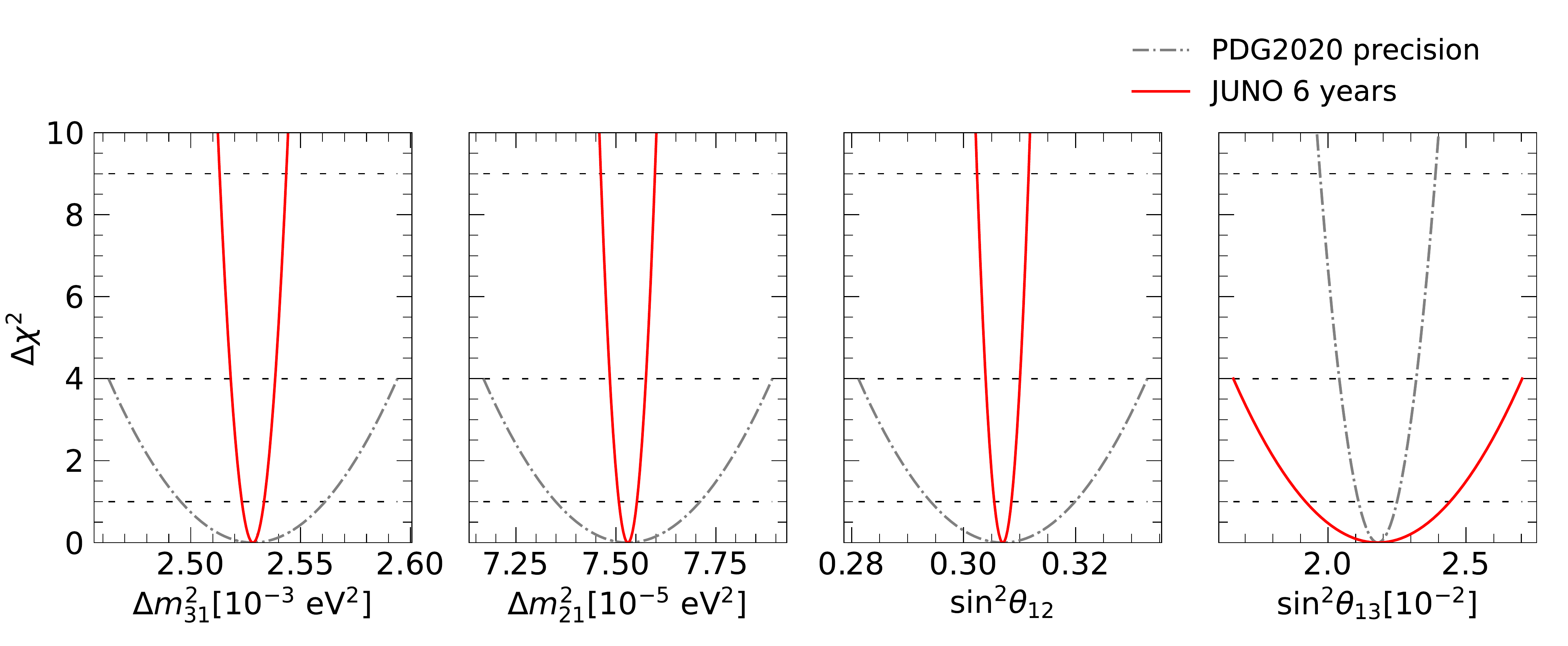}
  \caption{Comparison of 1-d $\Delta\chi^2$ distributions of oscillation parameters: Today (PDG2020, dashed curve) v.s. projection with 6~years data taking of JUNO (solid red curve)}
  \label{fig:sensitivity}
\end{figure}

\begin{figure}[!h]
  \centering
  \includegraphics[width=0.9\textwidth]{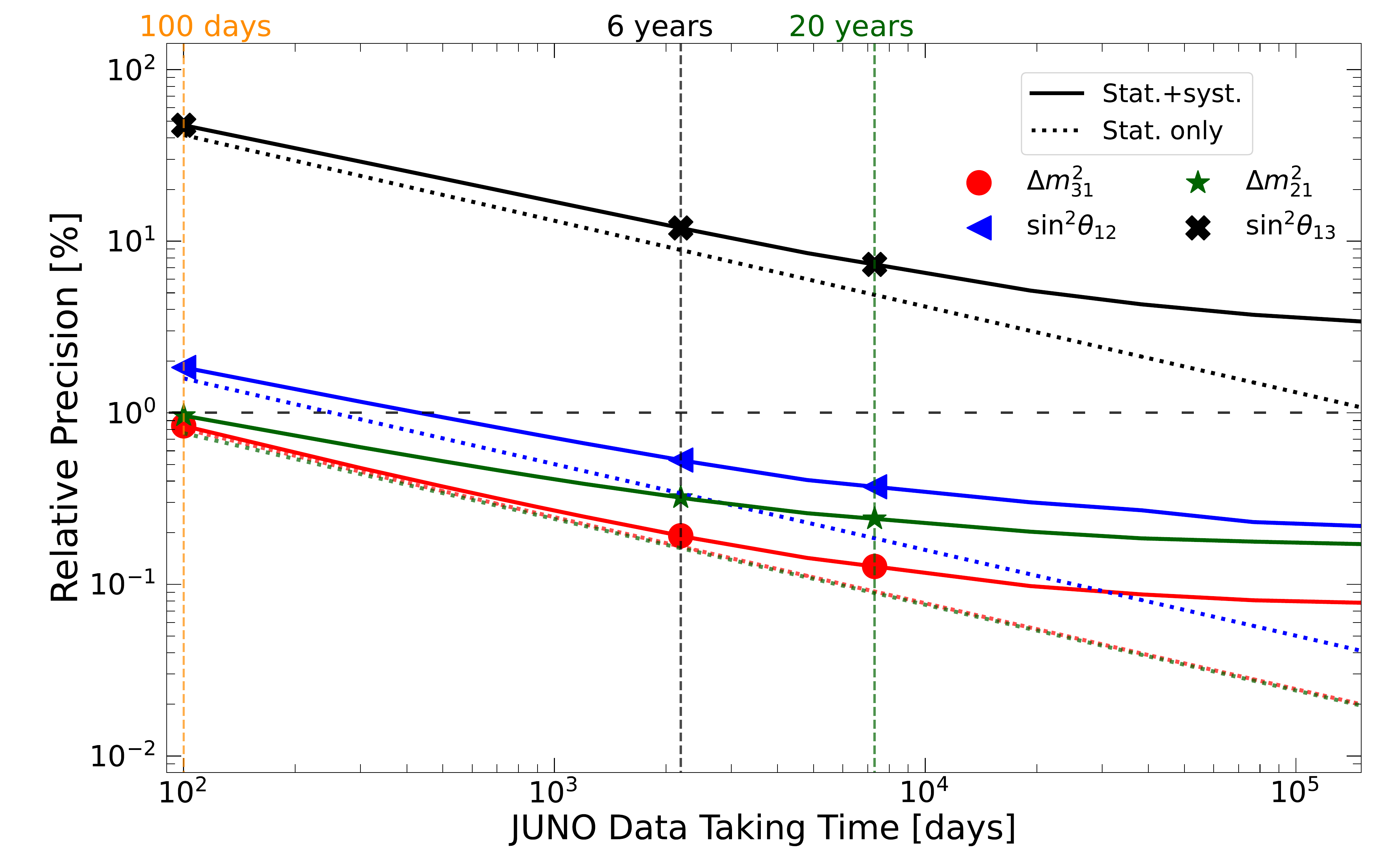}
  \caption{Relative precision of the oscillation parameters as a function of JUNO data taking time. The markers and vertical lines stand for 100 days, 6 years, and 20 years of data taking. The horizontal gray dashed line stands for 1\% relative precision. The green dotted and red dotted lines are on top of each other since the statistical-only precision is essentially identical for the $\Delta m^2_{31}$ and $\Delta m^2_{21}$ parameters.
  }
  \label{fig:sensitivityT}
\end{figure}

The 1$\sigma$ uncertainty for $\Delta m^2_{31}$, $\Delta m^2_{21}$, $\sin^2 \theta_{12}$, and $\sin^2 \theta_{13}$ is calculated with all rate and shape systematic uncertainties in three different regimes of data-taking time: 100 days (statistics-dominated regime); 6 years (nominal); and 20 years (systematics-dominated regime). The 1$\sigma$ limits of each parameter are obtained by marginalizing over all others, and finding the values for which $\Delta \chi^2$ changes by a unit.
The total precision obtained is summarized in Table~\ref{tab:sensitivity}. Additionally, Fig.~\ref{fig:sensitivity} shows the $\Delta\chi^2$ profiles of JUNO compared to today's state of the art knowledge~\cite{Zyla:2020zbs}. As shown there, JUNO is expected to improve upon today's precision by almost one order of magnitude for three out of six neutrino oscillation parameters, measuring them to the per mille precision.
In fact, about 100~days of data taking would be enough for JUNO to dominate the world precision on those parameters, although additional improvements are expected with more statistics. This is particularly the case for $\Delta m^2_{31}$, as coarsely quantified in Table~\ref{tab:sensitivity}, but fully illustrated in Fig.~\ref{fig:sensitivityT} where the impact of the systematic uncertainties can be observed via the deviation of the total sensitivity from the statistics-only limit.
\begin{table}[!h]
    \centering
    \resizebox{\columnwidth}{!}{%
    \begin{tabular}{l|ccccc}
    \hline
                                    &Central Value           & PDG2020          & 100\,days & 6\,years & 20\,years\\
    \hline
	$\Delta m^2_{31}$ ($\times 10^{-3}$ eV$^2$) & 2.5283      & $\pm0.034$ (1.3\%) & $\pm0.021$ (0.8\%) & $\pm0.0047$ (0.2\%) & $\pm0.0029$ (0.1\%)\\
	$\Delta m^2_{21}$ ($\times 10^{-5}$ eV$^2$) & 7.53      & $\pm0.18$ (2.4\%) & $\pm 0.074$ (1.0\%) & $\pm0.024$ (0.3\%) & $\pm0.017$ (0.2\%)\\
	$\sin^2 \theta_{12}$ 	                    & 0.307     & $\pm0.013$ (4.2\%) & $\pm0.0058$ (1.9\%) & $\pm0.0016$ (0.5\%) & $\pm0.0010$ (0.3\%)\\
	$\sin^2 \theta_{13}$                        & 0.0218    & $\pm0.0007$ (3.2\%) & $\pm0.010$ (47.9\%) & $\pm0.0026$ (12.1\%) & $\pm0.0016$ (7.3\%)\\
    \hline
    \end{tabular}%
    }
    \caption{A summary of precision levels for the oscillation parameters. The current knowledge (PDG2020~\cite{Zyla:2020zbs}) is compared with 100~days, 6~years, and 20~years of JUNO data taking. No external constraint on $\sin^2 \theta_{13}$ is applied for these results.}
    \label{tab:sensitivity}
\end{table}%

\begin{figure}[!h]
  \centering
  \includegraphics[width=0.45\textwidth]{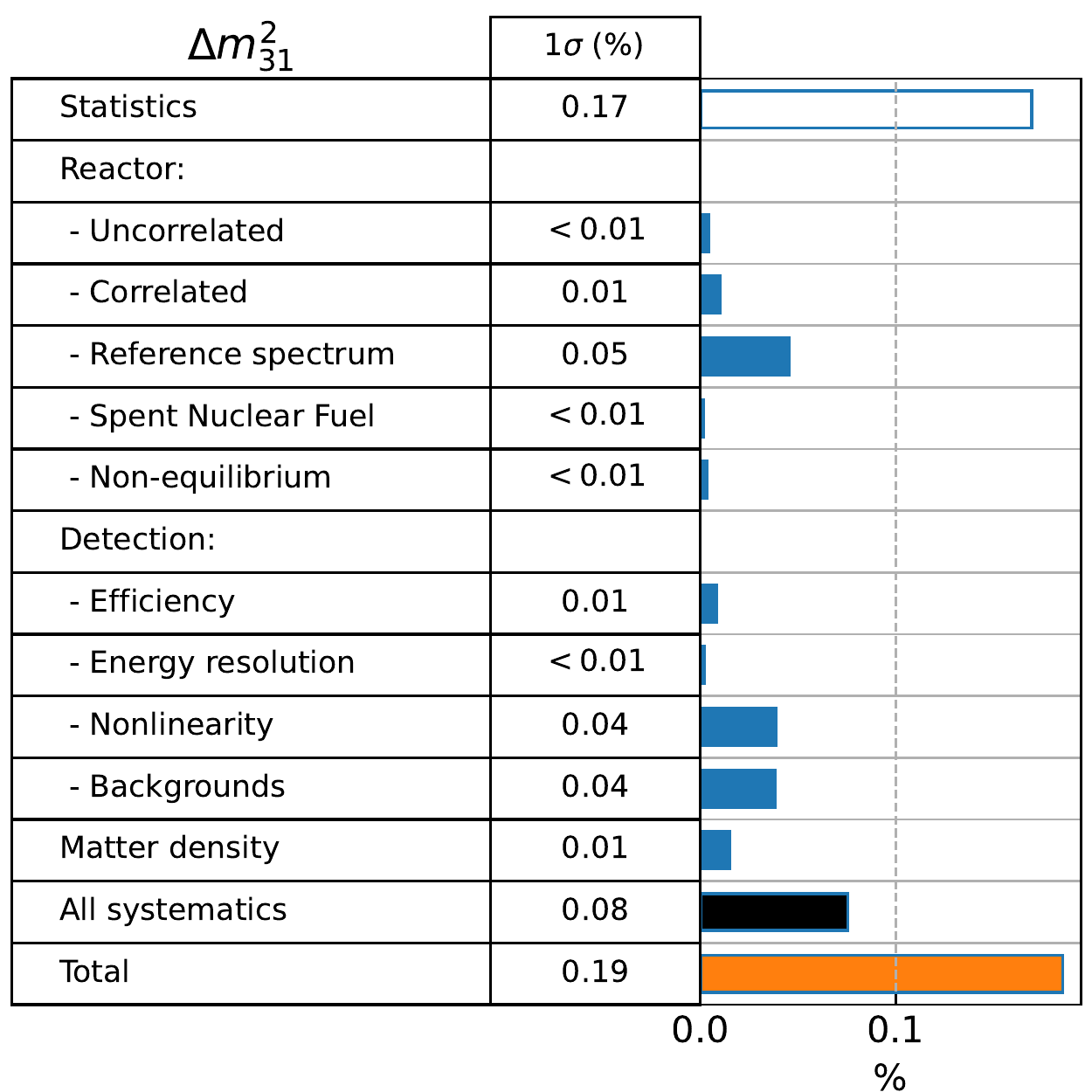}
  \includegraphics[width=0.45\textwidth]{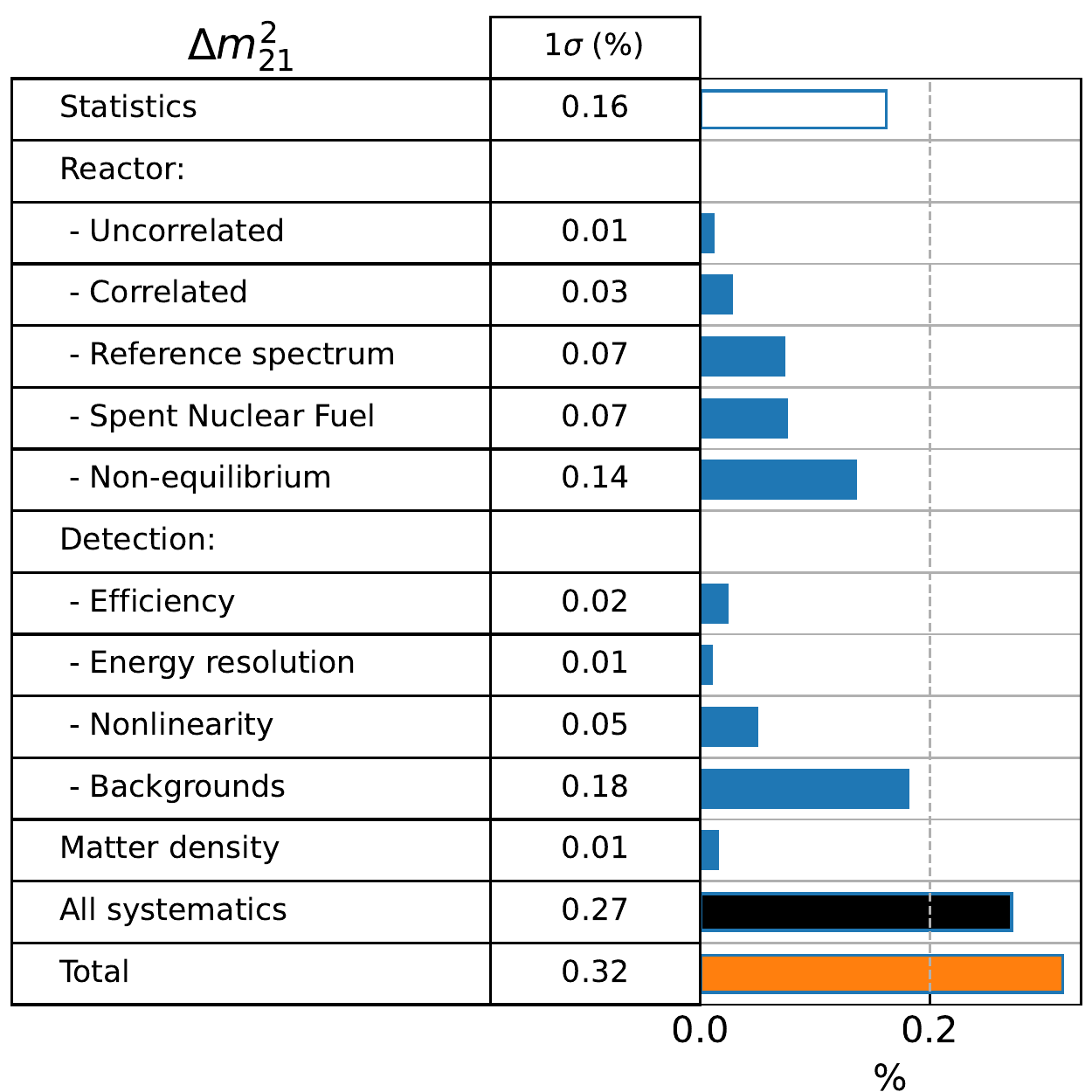}
  \includegraphics[width=0.45\textwidth]{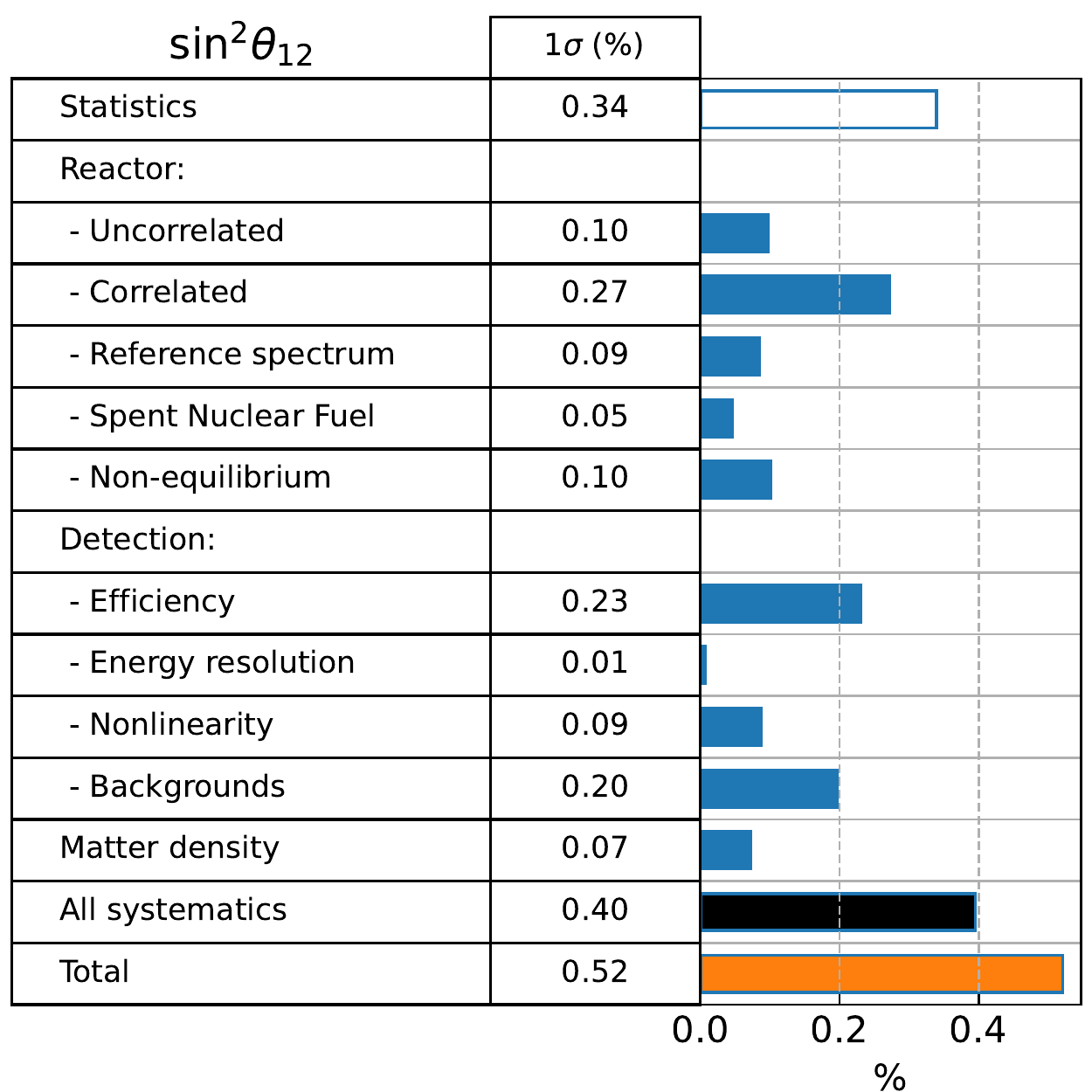}
  \includegraphics[width=0.45\textwidth]{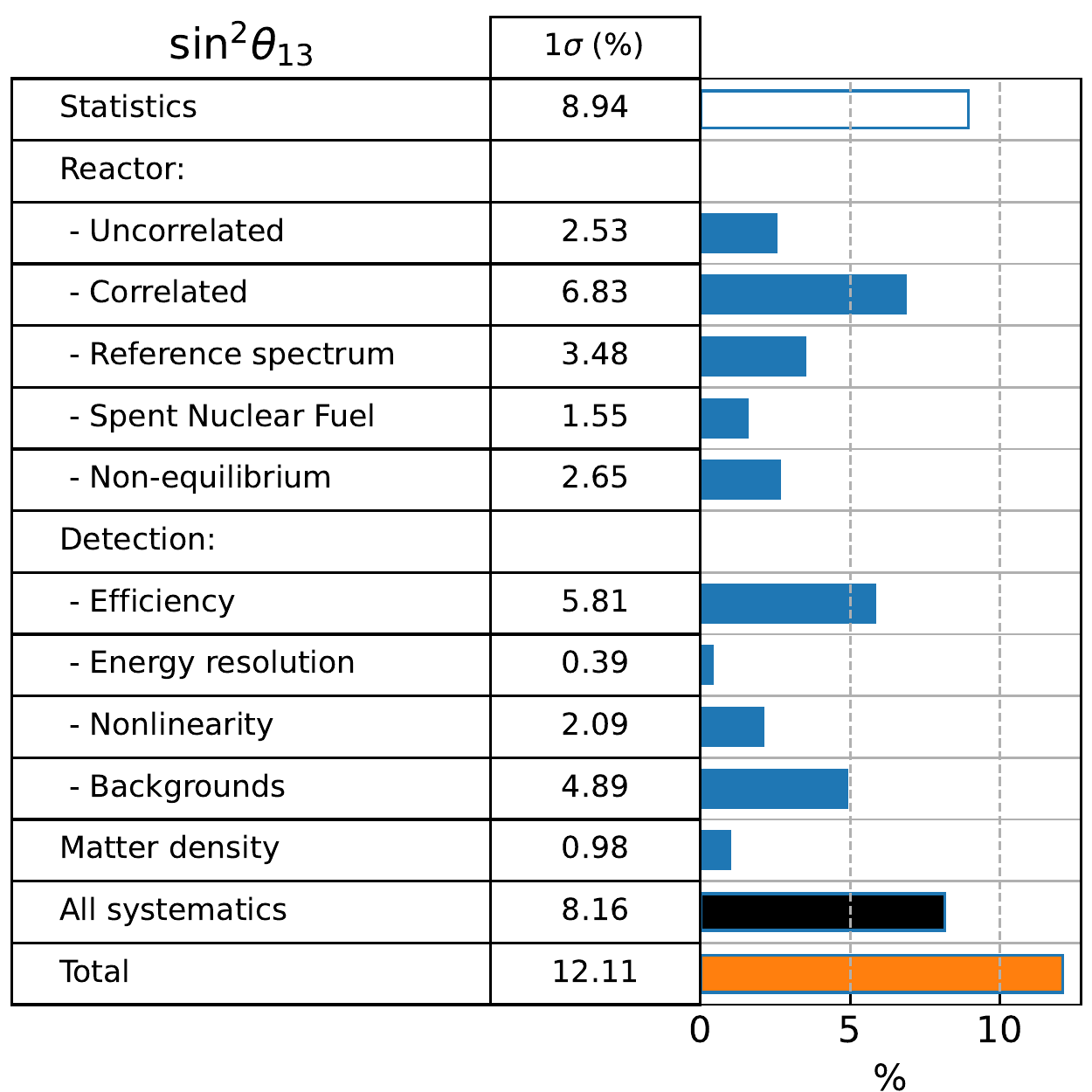}
  \caption{
  An illustration of the relative impact of individual sources of uncertainty on the total precision of the $\Delta m^2_{31}$, $\Delta m^2_{21}$, $\sin^2\theta_{12}$, and $\sin^2\theta_{13}$ oscillation parameters. The empty boxes represent the uncertainty resulting from considering only the statistical uncertainty of the reactor antineutrino sample. The impact of each source of systematic error, represented by the filled blue boxes, is assessed by enabling the corresponding uncertainty together with the statistical uncertainty and removing the latter. The removal is done by assuming that the statistical and systematic uncertainties add in quadrature, which allows to isolate the systematic component as $\sqrt{ (stat.+syst.)^{2} - (stat.)^{2}}$. The filled black box on every graph is obtained using the same procedure but simultaneously turning on all sources of systematic uncertainty rather than one at a time.
  The total uncertainty resulting from simultaneously considering all sources of statistical and systematic error is shown in the last orange row of each table. All uncertainties correspond to six years of JUNO data and are reported as relative uncertainty contributions to the precision of the particular oscillation parameter.
  }
  \label{fig:sensitivity_breakdown}
\end{figure}

The breakdown of statistical and systematic uncertainties on each parameter is shown in Fig.~\ref{fig:sensitivity_breakdown} for a nominal exposure of 6 years, allowing to identify the most important systematic effects.
The statistics-only sensitivity is also provided so that the relative impact of the systematic uncertainties can be easily seen.

The sensitivity of the two mixing angles is dominated by rate systematic uncertainties, mainly from the reactor flux normalization and the detector efficiency, both of which affect the analysis in the exact same way. Rate uncertainties have dominated most measurements of oscillation parameters to date, but their impact here is mitigated by JUNO's rich spectral shape information that provides a constraint on the normalization. Even though the reactor correlated uncertainty (due to the mean cross-section per fission and energy per fission uncertainties) is roughly double the efficiency uncertainty as indicated in Table~\ref{tab:ratesyst}, their impact on the mixing angles is quite similar as seen in Fig.~\ref{fig:sensitivity_breakdown}, differing by $\sim$15\% relative. In fact, if these uncertainties are increased very significantly, their impact on the precision changes very little from what it is shown. As a matter of fact, JUNO will be the first oscillation experiment where the spectral information provides such a good constraint on the normalization.

On the other hand, the sensitivity to the two mass splittings is dominated by systematic uncertainties distorting the spectral shape, mainly those from the reference spectrum and the detector nonlinearity. The spent nuclear fuel, non-equilibrium, and background systematic effects also distort the spectral shape, particularly in the low energy region, impacting the precision of $\Delta m^2_{21}$. The precision of the $\Delta m^2_{31}$ and $\sin^2 \theta_{13}$ parameters is statistics dominated even after six years of data taking, as indicated in Fig.~\ref{fig:sensitivityT}.

Figure~\ref{fig:sensitivity_breakdown} also shows that the impact of the density uncertainty in the calculation of the matter effects is small. Nevertheless, it is important to consider this effect when fitting the oscillation parameters, since it impacts the central value of the measurement, as discussed in the beginning of Section~\ref{sec:NO_Matter}.

Table~\ref{tab:sensitivity} and Fig.~\ref{fig:sensitivityT} show that JUNO alone has very limited ability to constrain $\sin^2 \theta_{13}$ beyond today's world knowledge. On the other hand, if we constrain $\sin^2 \theta_{13}$ with the uncertainty from PDG2020~\cite{Zyla:2020zbs}, the relative improvement in the precision of the other three parameters is smaller than 0.3\% with 6 years of data.

The impact of the neutrino mass ordering choice on the sensitivity of the parameters was also evaluated and found to be negligible. Therefore, the nominal results presented here are good for both the normal and inverted ordering hypotheses. Using the wrong ordering (e.g. using inverted ordering in a sample where normal ordering was assumed) produced sensitivities that are no larger than 5\% of the nominal values.

As discussed in Section~\ref{sec:JUNO_Reactor}, our nominal analysis considered the neutrino flux from the eight reactors $\sim$52.5~km away from the JUNO detector, the six in the Daya Bay power plant and, as a background, all the other reactors in the world. The flux from the Huizhou power plant was not included since no official date for its start of operations was available at the time of writing. However, assuming that the Huizhou power plant is operational from the beginning of JUNO data-taking degrades the sensitivities by less than a relative 3\%.

Section~\ref{sec:detector} describes how JUNO has two independent PMT systems (SPMT and LPMT) for photo-detection with different photon occupancy regimes. The ability to perform a measurement of the oscillation parameters using only the SPMT system ($\sigma_{E}\sim$17\% for 1 MeV energy depositions) was also evaluated. While the measurement of the fast oscillation in Fig.~\ref{fig:spectrum} driven by $\sin^22\theta_{13}$ and $\Delta m^2_{31}$ requires a very good energy resolution, the solar parameters $\sin^22\theta_{12}$ and $\Delta m^2_{21}$ that drive the low frequency oscillation can be measured with the SPMT system alone (see Fig.~\ref{fig:response}). Using the simple model of Section~\ref{subsec:det_res} to describe the energy resolution of this system, it was found that the expected precision of the two solar parameters is only less than 5\% worse than the nominal results using the combined LPMT+SPMT system. This will provide a valuable internal validation of these parameters' measurement with some different systematic uncertainties, namely those uncorrelated across the two systems.

\begin{figure}[!h]
  \centering
  \includegraphics[width=0.9\textwidth]{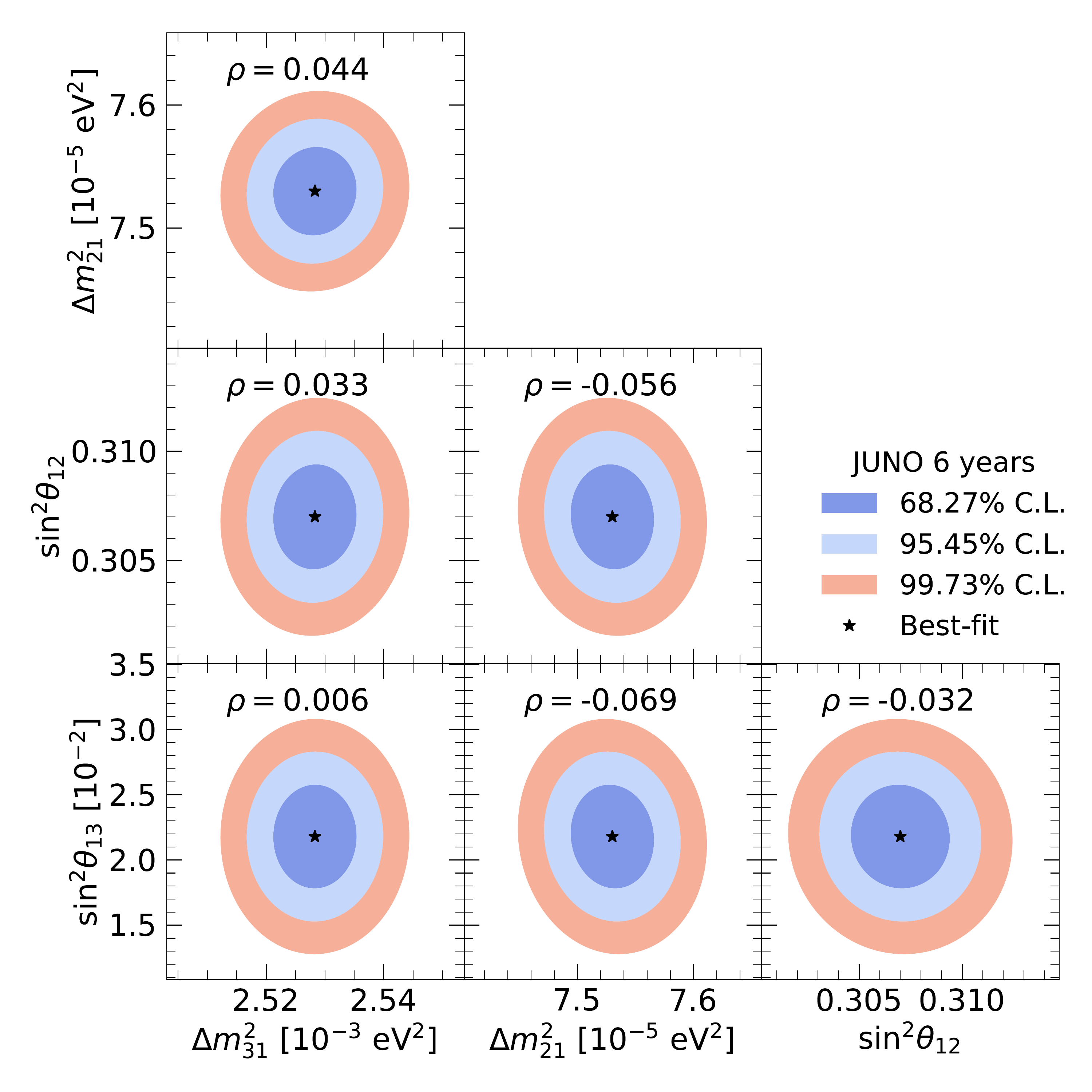}
  \caption{Two-dimensional 68.27\%, 95.45\%, and 99.73\% confidence level (C.L.) contours for all pairs of oscillation parameters that JUNO is sensitive to. These contours correspond to the 1, 2, and 3$\sigma$ confidence regions, respectively. For each point in these projections, the underlying $\Delta \chi^2$ value is obtained by minimizing over the other two parameters. The black stars represent the best-fit points as well as the true values of the oscillation parameters. The correlation coefficient between each pair of oscillation parameters is shown as $\rho$.
  }
  \label{fig:sensitivityC}
\end{figure}

Finally, the stability and precision of the results are illustrated by the correlations between the oscillation parameters shown in Fig.~\ref{fig:sensitivityC}. This figure depicts how these parameters are nearly uncorrelated, highlighting the abundant information available in JUNO's high-resolution measurement of the reactor antineutrino spectrum. It also explains the small impact of constraining $\sin^2 \theta_{13}$ as discussed above. Each parameter has a specific effect on the spectral shape that is retrievable with minimum interference between the parameters through the analysis described here.

\section{Conclusion}

JUNO is a next-generation liquid scintillator neutrino observatory under construction at a baseline of about $52.5$~km from eight nuclear reactors in the south of China. Its unprecedented size and energy resolution will enable it to make a precise measurement of the oscillated spectrum of reactor antineutrinos and to observe the so-called solar and atmospheric oscillation effects simultaneously for the first time. As a result, the $\Delta m^2_{31}$, $\Delta m^2_{21}$, and $\sin^2 \theta_{12}$ oscillation parameters will be determined to significantly better than sub-percent precision. Updated sensitivities to these parameters have been obtained using the most recent information available to date about the experimental site's location and overburden, the detector response, the backgrounds, the surrounding nuclear reactors, and the reactor antineutrino spectral shape constraints expected from the TAO satellite detector. The sensitivities were assessed with four independent analyses using alternative but equally rigorous treatments of the systematic uncertainties, all of which yielded results in excellent agreement with each other. With six years of JUNO data taking, the $\Delta m^2_{31}$, $\Delta m^2_{21}$, and $\sin^2 \theta_{12}$ parameters will be determined to a world-leading precision of $0.5\%$, $0.3\%$, and $0.2\%$, respectively. These measurements will constitute an important input to other experiments, provide constraints for model building, and enable more precise searches for physics beyond the Standard Model in the neutrino sector.

\input{JUNOacknowledgement}

\bibliographystyle{h-physrev5}
\bibliography{references}
\end{document}

%% file: fullname_211222.tex
\author[5,41]{Angel Abusleme}
\author[46]{Thomas Adam}
\author[67]{Shakeel Ahmad}
\author[67]{Rizwan Ahmed}
\author[56]{Sebastiano Aiello}
\author[67]{Muhammad Akram}
\author[67]{Abid Aleem}
\author[49]{Tsagkarakis Alexandros}
\author[29]{Fengpeng An}
\author[22]{Qi An}
% ...
\affil[5]{Pontificia Universidad Cat\'{o}lica de Chile, Santiago, Chile}
% ...
\affil[22]{University of Science and Technology of China, Hefei, China}
% ...
\affil[29]{East China University of Science and Technology, Shanghai, China}
% ...
\affil[41]{Millennium Institute for SubAtomic Physics at the High-energy Frontier (SAPHIR), Chile}
% ...
\affil[46]{IPHC, Universit\'{e} de Strasbourg, CNRS/IN2P3, F-67037 Strasbourg, France}
% ...
\affil[49]{III. Physikalisches Institut B, RWTH Aachen University, Aachen, Germany}
% ...
\affil[56]{INFN Catania and Dipartimento di Fisica e Astronomia dell Universit\`{a} di Catania, Catania, Italy}
% ...
\affil[67]{Pakistan Institute of Nuclear Science and Technology, Islamabad, Pakistan}
% ...

%% file: JUNOacknowledgement.tex
\section*{Acknowledgements}

We are grateful for the ongoing cooperation from the China General Nuclear Power Group.
This work was supported by
the Chinese Academy of Sciences,
the National Key R\&D Program of China,
the CAS Center for Excellence in Particle Physics,
Wuyi University,
and the Tsung-Dao Lee Institute of Shanghai Jiao Tong University in China,
the Institut National de Physique Nucl\'eaire et de Physique de Particules (IN2P3) in France,
the Istituto Nazionale di Fisica Nucleare (INFN) in Italy,
the Italian-Chinese collaborative research program MAECI-NSFC,
the Fond de la Recherche Scientifique (F.R.S-FNRS) and FWO under the ``Excellence of Science – EOS” in Belgium,
the Conselho Nacional de Desenvolvimento Cient\'ifico e Tecnol\`ogico in Brazil,
the Agencia Nacional de Investigacion y Desarrollo and ANID - Millennium Science Initiative Program - ICN2019\_044 in Chile,
the Charles University Research Centre and the Ministry of Education, Youth, and Sports in Czech Republic,
the Deutsche Forschungsgemeinschaft (DFG), the Helmholtz Association, and the Cluster of Excellence PRISMA+ in Germany,
the Joint Institute of Nuclear Research (JINR) and Lomonosov Moscow State University in Russia,
the joint Russian Science Foundation (RSF) and National Natural Science Foundation of China (NSFC) research program,
the MOST and MOE in Taiwan,
the Chulalongkorn University and Suranaree University of Technology in Thailand,
University of California at Irvine and the National Science Foundation in USA.